\documentclass[pra,reprint,superscriptaddress,twocolumn]{revtex4-2}
\usepackage[utf8]{inputenc}
\usepackage{adjustbox}
\usepackage{amsfonts}
\usepackage{amsmath}
\usepackage{amsthm}
\usepackage{amssymb}
\usepackage{braket}
\usepackage{graphicx}
\usepackage{subcaption}
\usepackage{svg}
\usepackage{xcolor}
\usepackage[font=small,labelfont=bf,justification=justified]{caption}
\usepackage{lipsum}

\usepackage{tikz}
\usetikzlibrary{shapes.geometric, arrows}

\tikzstyle{blank} = [rectangle, rounded corners, minimum width=2cm, minimum height=0.75cm,text centered]
\tikzstyle{red} = [rectangle, rounded corners, minimum width=2cm, minimum height=0.75cm,text centered, draw=black, fill=red!30]
\tikzstyle{teal} = [rectangle, rounded corners, minimum width=2cm, minimum height=0.75cm,text centered, draw=black, fill=teal!30]
\tikzstyle{violet} = [rectangle, rounded corners, minimum width=2cm, minimum height=0.75cm,text centered, draw=black, fill=violet!20]
\tikzstyle{yellow} = [rectangle, rounded corners, minimum width=2cm, minimum height=0.75cm,text centered, draw=black, fill=yellow!20]
\tikzstyle{blue} = [rectangle, minimum width=2cm, minimum height=0.75cm, text centered, draw=black, fill=blue!30]
\tikzstyle{orange} = [rectangle, minimum width=2cm, minimum height=0.75cm, text centered, draw=black, fill=orange!30]
\tikzstyle{cyan} = [rectangle, minimum width=2cm, minimum height=0.75cm, text centered, draw=black, fill=cyan!30]
\tikzstyle{gray} = [rectangle, minimum width=2cm, minimum height=0.75cm, text centered, draw=black, fill=gray!30]
\tikzstyle{arrow} = [thick,->,>=stealth]
\tikzstyle{arrowdotted} = [dotted,->,>=stealth]

\newcommand{\pars}[1]{\mathopen{}\left(#1\right)\mathclose{}}

\newcommand{\braces}[1]{\mathopen{}\left\{#1\right\}\mathclose{}}

\newcommand{\sumbinom}[2]{\binom{#1}{#2}_{\sum}}

\begin{document}

\title{A fast and frugal Gaussian Boson Sampling emulator}

\author{Tom Dodd}
\affiliation{School of Informatics, University of Edinburgh, 10 Crichton Street, Edinburgh EH8 9AB, United Kingdom}

\author{Javier Mart\'{i}nez-Cifuentes}
\affiliation{D\'epartement de g\'enie physique, \'Ecole polytechnique de Montr\'eal, Montr\'eal, QC, H3T 1J4, Canada}

\author{Oliver Thomson Brown}
\affiliation{EPCC, University of Edinburgh, 47 Potterow, Edinburgh EH8 9BT, United Kingdom}

\author{Nicol\'as Quesada}
\affiliation{D\'epartement de g\'enie physique, \'Ecole polytechnique de Montr\'eal, Montr\'eal, QC, H3T 1J4, Canada}

\author{Raúl García-Patrón}
\affiliation{School of Informatics, University of Edinburgh, 10 Crichton Street, Edinburgh EH8 9AB, United Kingdom}
\affiliation{Phasecraft Ltd., 77-79 Charlotte Street, London W1T 4PW, United Kingdom}

\begin{abstract}
If classical algorithms have been successful in reproducing the estimation of expectation values of observables of some quantum circuits using off-the-shelf computing resources, matching the performance of the most advanced quantum devices on sampling problems usually requires extreme cost in terms of memory and computing operations, making them accessible to only a handful of supercomputers around the world. In this work, we demonstrate for the first time a classical simulation outperforming Gaussian boson sampling experiments of one hundred modes on established benchmark tests using a single CPU or GPU. Being embarrassingly parallelizable, a small number of CPUs or GPUs allows us to match previous sampling rates that required more than one hundred GPUs. We believe algorithmic and implementation improvements will generalize our tools to photo-counting, single-photon inputs, and pseudo-photon-number-resolving scenarios beyond one thousand modes. Finally, most of the innovations in our tools remain valid for generic probability distributions over binary variables, rendering it potentially applicable to the simulation of qubit-based sampling problems and creating classical surrogates for classical-quantum algorithms.
\end{abstract}

\maketitle

The last six years have seen an increasing number of implementations of quantum algorithms claiming to have reached quantum computational advantage over their classical counterparts; these range from random circuit sampling on quantum superconducting devices \cite{Arute_2019,gao2025establishing}, Gaussian or single-photon boson sampling on photonic hardware~\cite{zhong2020quantum,zhong2021phase,madsen2022quantum,deng2023gaussian,liu2025} and atomic ensembles ~\cite{young2024atomic}, respectively, or the estimation of dynamical observables in trapped ions ~\cite{25Haghshenas} and superconducting circuits~\cite{Kim2023}. On hypothetical error-free quantum hardware, all the candidate proposals are believed to be out of reach for classical computers, challenging today's most advanced supercomputers with sizes beyond 50 qubits or hundreds of modes and photons. For sampling problems, this is often backed by complexity-theoretic arguments~\cite{bouland2019complexity,aaronson2011computational,deshpande2022quantum,grier2022complexity,bouland2023complexity}. 
\begin{figure*}[t]
    \centering
    \includegraphics[width=0.8\textwidth]{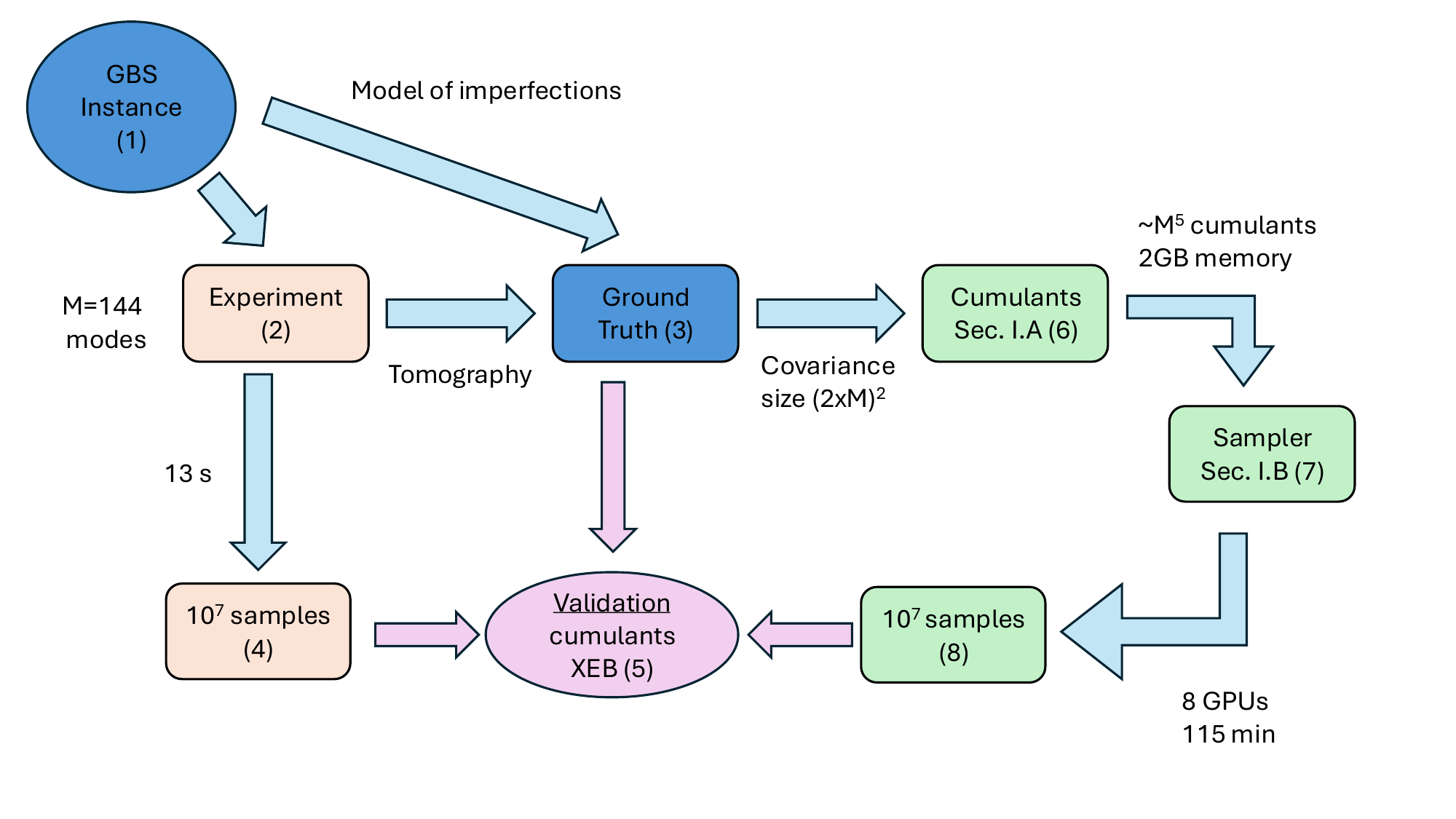}
    \caption{(1) An instance of GBS is characterized by its ideal optical circuit, or equivalently the covariance matrix of its associated Gaussian state; (2) An experiment implements a version affected by losses and other imperfections; (3) One can extract the ground truth of the experiment from tomographic data or applying a good model of imperfections to the noiseless GBS instance; (4) 10 million samples are generated by quantum hardware in approximately 10 seconds; (5) The samples are benchmarked using cumulant statistical analysis and XEB; (6) The first stage of our algorithm computes a list of $\binom{144}{5}$ correlators that are later transformed into cumulants from the ground truth covariance matrix, as detailed in section \ref{subsec:Training}; (7) The second stage generates samples from the corresponding cumulants as detailed in section \ref{subsec:Algorithm}; (8) Using 8 GPUs, we can generate 10 million samples in 115 minutes, later validated with the same benchmarking tests.} 
    \label{fig:concept}
\end{figure*}

In reality, current quantum devices are prone to errors and imperfections that limit their performance. Unless fault-tolerance based on quantum error-correction is implemented, the quality of the computations decays rapidly. The pervasive presence of errors is the key enabler behind all classical algorithm implementations that claim to have successfully disproved previous claims of quantum advantage, such as Refs.~\cite{pan2022simulation,pan2022solving} for random quantum circuits, Ref.~\cite{oh2024classical} for Gaussian boson sampling (GBS), and Ref.~\cite{kechedzhi2024effective,rudolph2023,begusic2023} for the estimation of observables of complex spin-system dynamics. 

If classical algorithms have been successful in reproducing the estimation of expectation values of observables of some recent quantum experiments, even using off-the-shelf computing resources~\cite{kechedzhi2024effective,rudolph2023,begusic2023}, the simulation of sampling tasks, both qubit- and bosonic-based, are known to pose significantly harder challenges to classical simulations. The implementations of these algorithms required an extreme cost in terms of memory and computing operations, making them accessible to only a handful of supercomputers around the world and unreasonable to reproduce on a regular basis.  
To date, there has been no demonstration of a classical algorithm that matches the performance on standard benchmarking tests of the most advanced quantum devices on a sampling problem using off-the-shelf computational resources. 

In this work we demonstrate for the first time a classical simulation of imperfect GBS matching and sometimes outperforming the actual quantum experiments in a series of standardized tests, with the need of only a single CPU or GPU, while Ref.~\cite{oh2024classical} required more than one hundred GPUs.
Additionally, our implementation for 144 modes requires 2 GB to encode the quantum state before the measurement, three orders of magnitude less than in Ref.~\cite{oh2024classical}. Being embarrassingly parallelizable, we were able to generate 10 million samples in 115 minutes using 8 contemporary GPUs on one single cluster node. These significant improvements in memory use, running time and energy costs makes mainstream what was before accessible only to a handful of facilities in the world.

Our main technical contribution is the design of an algorithm
to sample from the joint probabilities of $M$ bit samples by exploiting a decomposition in terms of cumulants of increasing order and approximated marginals of the original distribution. Truncating the expression to order $K$ allows us to capture all relevant statistics of the hardware experiment while making its running time scale as $O(M^K)$.

The algorithm benefiting from a polynomial scaling of order $K$, should remain a contender against future hardware improvements and larger system sizes, such as the recent Jiuzhang 4.0 experiments \cite{liu2025}. We believe the algorithm is easily generalizable to photo-counting and single-photon scenarios. 
Finally, most of the innovations in our tools remain valid for generic probability distributions over binary variables, rendering it applicable to the simulation of qubit-based sampling problems and creating classical surrogates for hybrid classical-quantum algorithms.

\section{Emulator description}\label{sec:new emulator}

Error-free GBS is believed to be a hard sampling problem where an instance over $M$ modes is defined by a set of squeezed state sources located at a given set of input ports and a linear-optical interferometer, characterized by an $M\times M$ unitary matrix $U$ selected randomly over the Haar measure. Except for the Borealis experiment~\cite{madsen2022quantum}, where photo-counting was implemented at the output, all other GBS demonstrations so far have implemented threshold detection ~\cite{zhong2020quantum,zhong2021phase,deng2023gaussian}. In this work we focus on the second type of GBS, consisting of binary outcomes encoding the absence or presence of photons.

Historically, the ground truth of the GBS experiment used in its validation tests is the description of the Gaussian state at the output of the real interferometer taking into account the presence of losses. This approach is different from the one used in, for example, random circuit sampling, where one computes benchmarks such as the cross linear entropy with respect to the ideal noiseless instance. Whether it is more appropriate to use a lossy ground truth for photonic quantum advantage remains an open research problem~\cite{martinez2024linear}. With our goal being to design an emulator of noisy GBS, it is natural and non-controversial for us to consider the lossy ground truth as the benchmark to compare. 

As described in Figure \ref{fig:concept}, the ground truth is obtained from a 
characterization of the GBS device's transmission matrix and squeezing parameters. Alternatively, one could envision constructing the ground truth from the ideal GBS instance combined with an accurate model of imperfections.
Figure \ref{fig:concept} summarizes our emulator as a two-step process. We start from the ground truth Gaussian state's $(2M\times 2M)$ covariance matrix and compute $O(M^K)$ Fourier coefficients up to order $K$ and the corresponding cumulants, following the methodology explained in subsection \ref{subsec:Training}. The choice of the truncation order $K$ was made heuristically from a study of the experimental data, though one could develop and exploit analytical tools bounding the quality of the approximation.
as a function of $K$. 

The second step is a sampling algorithm described in subsection \ref{subsec:Algorithm} which generates samples using a chain rule of conditional probabilities, where each new bit is chosen according to the bias between 0 and 1 that is computed from a large weighted sum composed of pre-computed cumulants and approximations of marginals computed following a dynamical programming approach.

The certification of large-scale GBS is often performed using a number of different metrics. Firstly, Bayesian and heavy output generation (HOG) tests have been used to compare the experimental data against different hypotheses or spoofers/simulators, respectively. More robust approaches are the comparison of the click cumulants and a variant of cross-entropy benchmarking (XEB) that focuses on a post-selected set of data within a range of total photon counts, which have been used to provide evidence of quantum advantage by hardware implementations~\cite{zhong2020quantum,zhong2021phase,madsen2022quantum,deng2023gaussian}. Therefore, we will follow the recent classical algorithm in \cite{oh2024classical}, and use cross-linear entropy and cumulant distributions to benchmark our emulator.

\subsection{Pre-computing the cumulants}
\label{subsec:Training}

A cumulant $\kappa(S)$ of order $d$ captures all ``novel'' correlations present in a $d$-mode marginal distribution that are not captured by its lower-order cumulants of orders 1 through $d-1$. This property is nicely captured by the recursive definition of a cumulant for a set of random variables $S=\{i_1,\ldots,i_{d}\}$, where $i_j\in[M]\equiv\{1,\dots, M\}$:
\begin{equation}
    \kappa(S)= c(S) - \sum_{\pi \in \mathcal{P}(S)\setminus \{S\}} \prod_{B \in \pi} \kappa(B),
\end{equation}
where $\mathcal{P}(S)$ is the set of partitions of $S$, 
$\chi(A)=(-1)^{\sum_{k\in A}x_k}$ is the parity function over the set of variables $A$
and $c(A)=\sum_{\bar{x}_A\in\{0,1\}^{|A|}}\chi_A(\bar{x})\,p(\bar{x}_A)$ is its associated correlator. Note that $c(A)$ is known as a spin-correlator in many-body physics, and is, up to a factor of $1/2^{|A|}$, equivalent to the Fourier coefficient associated with the parity function $\chi(A)$ in the Fourier analysis of Boolean functions \cite{odonnell2021}.

The cumulants $\kappa(S)$ required for the sampling procedure are computed from the correlators $c(S)$ for all subsets $S$ of size 1 to $K$ of $[M]$ using the relation
\begin{equation}
\kappa(S)= \sum_{\pi \in \mathcal{P}(S)}(-1)^{|\pi|-1}(|\pi|-1)!\prod_{B\in \pi}c(B),
\label{eq:cumulant_corr_main}
\end{equation}
with $\mathcal{P}(S)$ the set of all partitions of $S$. The computation of the correlators is the only aspect of our algorithm which depends on the peculiarities of GBS. In the Appendix, we show how these correlators of order $d$ can be re-expressed analytically as a large sum of Gaussian integrals, leading to the expression
\begin{equation}
    c(S)= (-1)^{|S|}\sum_{R\subseteq S}\frac{(-2)^{|R|}}{\sqrt{\mathrm{det}\left[\frac{1}{\hbar}\left(\boldsymbol{\sigma}_R + \frac{\hbar}{2}\mathbb{I}\right)\right]}}.
    \label{eq:corre_cov_main}
\end{equation}
where $\boldsymbol{\sigma}_R$ is the $2R \times 2R$ covariance matrix of the subsystem $R$.
It is easy to see that the computation time for each correlator of order $d$ scales as $O(d^32^d)$ as it requires the computation of $2^d$ determinants of size at most $d$. Additionally, the computation time for each cumulant of order $d$ scales as $O(dB_d)$, where $B_d$ is the $d$th Bell number equalling the number of partitions of a set of size $d$, which compounds the complexity to $O(d(d/(e\ln{d}))^d)$. As we restrict to a small truncation order $K$, each term is extremely fast to compute, and the dominant factor in the scaling comes from the total number of correlators $\binom{M}{K}\approx O(M^K)$. 

\begin{figure*}[!ht]
    \centering
    \begin{tabular}{@{}c@{}}
         \begin{subfigure}[c]{0.45\linewidth}
            \includegraphics[width=\textwidth]{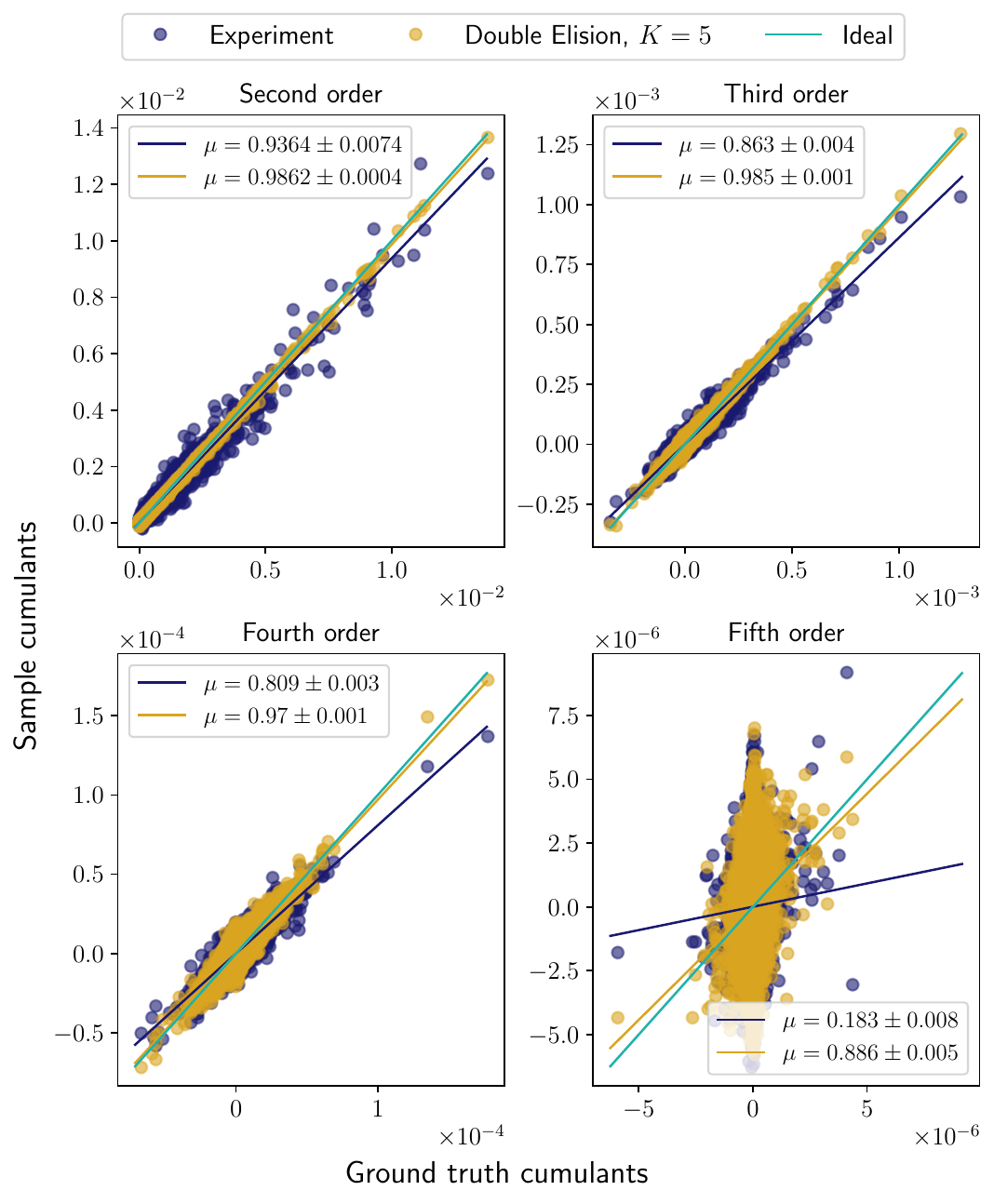}
         \end{subfigure}\\
         (a)
    \end{tabular}\qquad
    \begin{tabular}{@{}c@{}}
         \begin{subfigure}[c]{0.45\linewidth}
            \includegraphics[width=\textwidth]{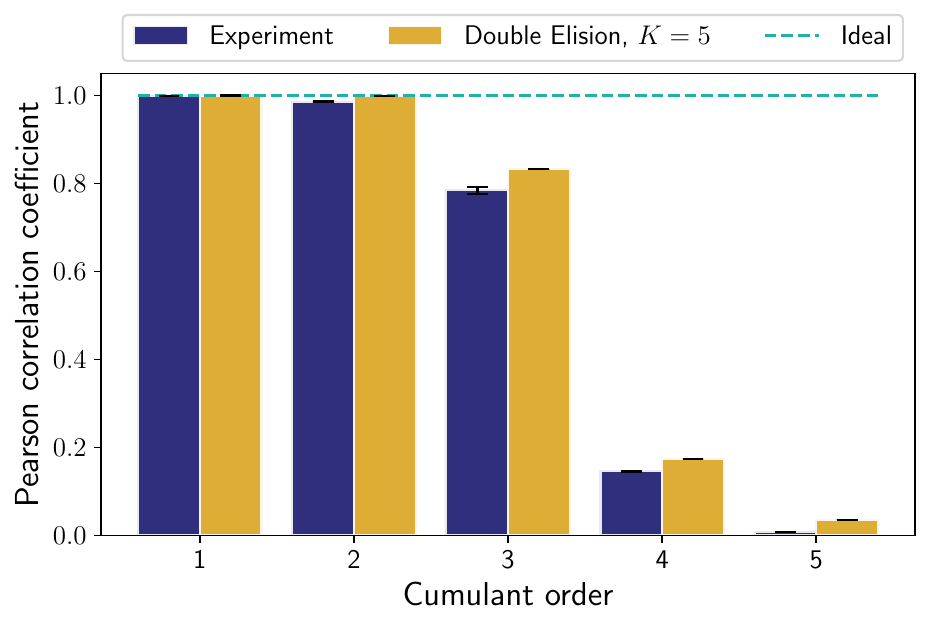}
         \end{subfigure}\\
         (b)\\
         \begin{subfigure}[c]{0.45\linewidth}
            \includegraphics[width=\textwidth]{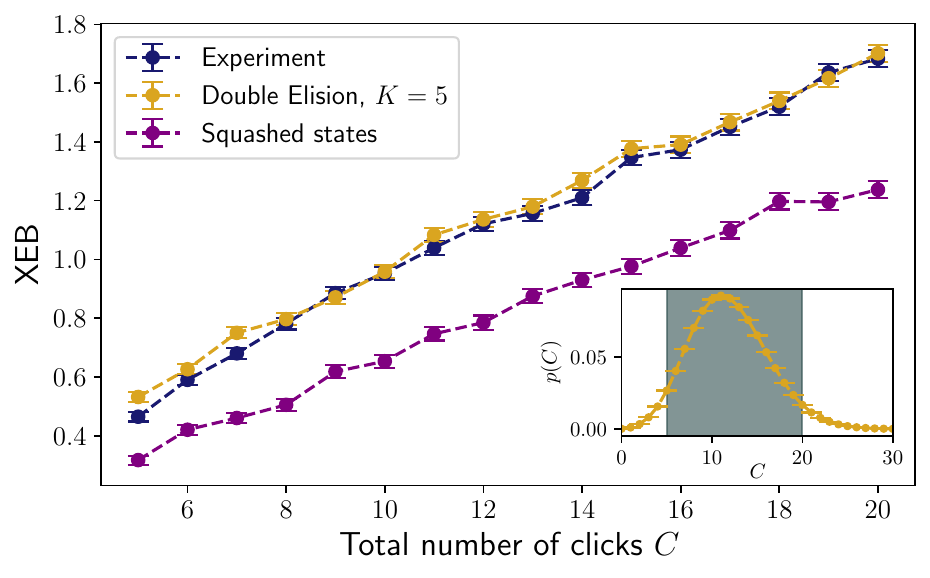}
         \end{subfigure}\\
         (c)\\
    \end{tabular}
    \caption{\textit{Validation tests for the J2-P65-2 experiment.} (a) Comparison between the estimated click cumulants (orders 2 to 5), from experimental (dark blue circles) and double-elision samples with $K=5$ (yellow circles), and those predicted by the ground truth of the experiment (light green line). The vertical axis corresponds to cumulants obtained from samples, either experimental or double elision, while the horizontal axis corresponds to the ground truth (i.e., theoretical) cumulants. The dark blue and yellow lines represent the linear fit between sample cumulants and those from the ground truth. The slope of each linear fit is shown in the legends. All cumulants were estimated using $10^7$ samples. (b) Pearson correlation coefficients between sample and ground truth cumulants. The light green dashed line indicates the ideal value of these coefficients, which is equal to 1. Error bars were obtained using 100 bootstrapping resamples. (c) XEB as a function of the total number of clicks $C$, for both experimental (dark blue dashed line with circles) and double-elision (yellow dashed line with circles) samples. For reference, we also include the XEB scores of the squashed states model~\cite{martinez2023classical} (purple dashed line with circles). Error bars were obtained through error propagation. The estimation of the XEB scores used $4000$ samples per number of clicks. The inset figure shows the total click distribution, $p(C)$, of the sampler. The shaded region indicates the range of total number of clicks, $C$, for which the XEB scores were computed.}
    \label{fig:validation_J2_2}
\end{figure*}

\subsection{The sampling algorithm}\label{subsec:Algorithm}
We are interested in sampling from a GBS distribution of $M$ binary outcomes.
The classical algorithm we construct is based on the chain rule of conditional probabilities $p(x_{n}|x_{n-1},\ldots,x_1)$.
Central to our algorithm is an expression for the bias $\Delta_{x_n} = p(0,x_{n-1},\ldots,x_0)-p(1,x_{n-1},\ldots,x_0)$ between the outputs 0 and 1 of the $n$th bit.  
The expression, presented in more detail in Appendix \ref{app:probs_cumul}, is a collection of terms containing cumulants of increasing order:
\begin{equation}\label{eq:samplecumulant}
    \Delta_{x_n} = \sum_{R\subseteq [n-1]}\frac{1}{2^{|R|}}\chi(R)\kappa(R\cup\{n\})\,p(\bar{R}),
\end{equation}
where $\kappa(R\cup\{n\})$ is the cumulant of order $|R|+1$ associated with the union of the singleton $\{n\}$ containing the index of the bit to be sampled and a subset $R$ of indices of bits sampled in the past, $\chi(R\cup\{n\})$ is its associated parity function, and $p(\bar{R})$ is the marginal probability over the complement of $R$ ($\bar{R}=[n-1]\setminus R$). Note that the term $R=\emptyset$ in the sum of~Eq.(\ref{eq:samplecumulant}) corresponds to the $n$th output being completely independent of all previous bits and fully determined by $\kappa(\{n\})=c(\{n\})$, while the subsequent terms in the sum add correlation with previous bits $x_1$ to $x_{n-1}$. From the value of $\Delta_{x_n}$ and the previously computed probability $p(x_{n-1},\ldots,x_1)$ 
one can sample the $n$th bit and compute $p(x_n,\ldots,x_1)$. 

Taking the sum over all partitions of $[n]$ into consideration would introduce a costly $O(2^n)$ scaling. 
It is often the case that local noise in large quantum circuits has the effect of strongly suppressing high-order correlations and cumulants.  
For example, an inspection of previous GBS experiments data shows that the value of cumulants beyond fifth order are statistically insignificant~\cite{zhong2020quantum,zhong2021phase,madsen2022quantum,deng2023gaussian}. This justifies truncating the sum in Eq.~(\ref{eq:samplecumulant}) to a constant order $K$ ($|R|+1\leq K$). The truncated version of $\Delta_{x_n}$ requires computing $\sum_{j=1}^K\binom{n}{j}\approx O(n^K)$ terms, each consisting of a cumulant, a parity function and a marginal $p(\bar{R})$.
With the dominant contribution coming from the last sampled bit, we have $O(M^K)$ terms 
in the truncated version of Eq.~(\ref{eq:samplecumulant}).

The $O(M^K)$ required cumulants have been computed in the pre-processing stage, using the definition presented in the previous subsection \ref{subsec:Training}. The parity functions are simple to compute with simple bitwise multiplication and additions. 
The only non-trivial part to compute are the marginals $p(\bar{R})$. Despite $\bar{R}$ matching previously appearing variables in the chain rule, $p(\bar{R})$ cannot be directly computed from previous data, as we would need to sum over variables $R$ for values that never occurred. In addition, the positions of these $R$ elided variables change with each term of the sum. This makes the exact calculation of $p(\bar{R})$ prohibitively expensive. The key aspect of constructing our algorithm is to design good approximations to the marginals $p(\bar{R})$. We present a pedagogical, simplified example in the Methods section and provide full detail in the Appendix \ref{App:marginals}. In short, our algorithm can be understood as a dynamical program where we keep track of a series of tables, one for each family of approximated marginals and one for the sampling probabilities $p(x_n,\ldots,x_1)$. At each step, the tables are updated from a non-trivial combination of previously computed values and pre-computed cumulants following specific rules inspired by an adaptation of Eq.~(\ref{eq:samplecumulant}) to marginal distributions. 

\section{Comparison to experiments}\label{sec:comparison}

In this section, we will compare the samples generated by our sampler together with the samples generated by different 144-mode Jiuzhang experiments, focusing on cases that were also analysed in Ref.~\cite{oh2024classical}. Throughout, we will use the convention established in Ref.~\cite{oh2024classical} for naming the different Jiuzhang configurations. All emulator samples were generated using the ``double-elision'' algorithm at order $K=5$, described in Appendix \ref{App:marginals}. In each case, we generated 10 million samples in 115 minutes using 8 GPUs on a cluster node.

\begin{figure*}[!ht]
    \centering
    \begin{tabular}{@{}c@{}}
         \begin{subfigure}[c]{0.45\linewidth}
            \includegraphics[width=\textwidth]{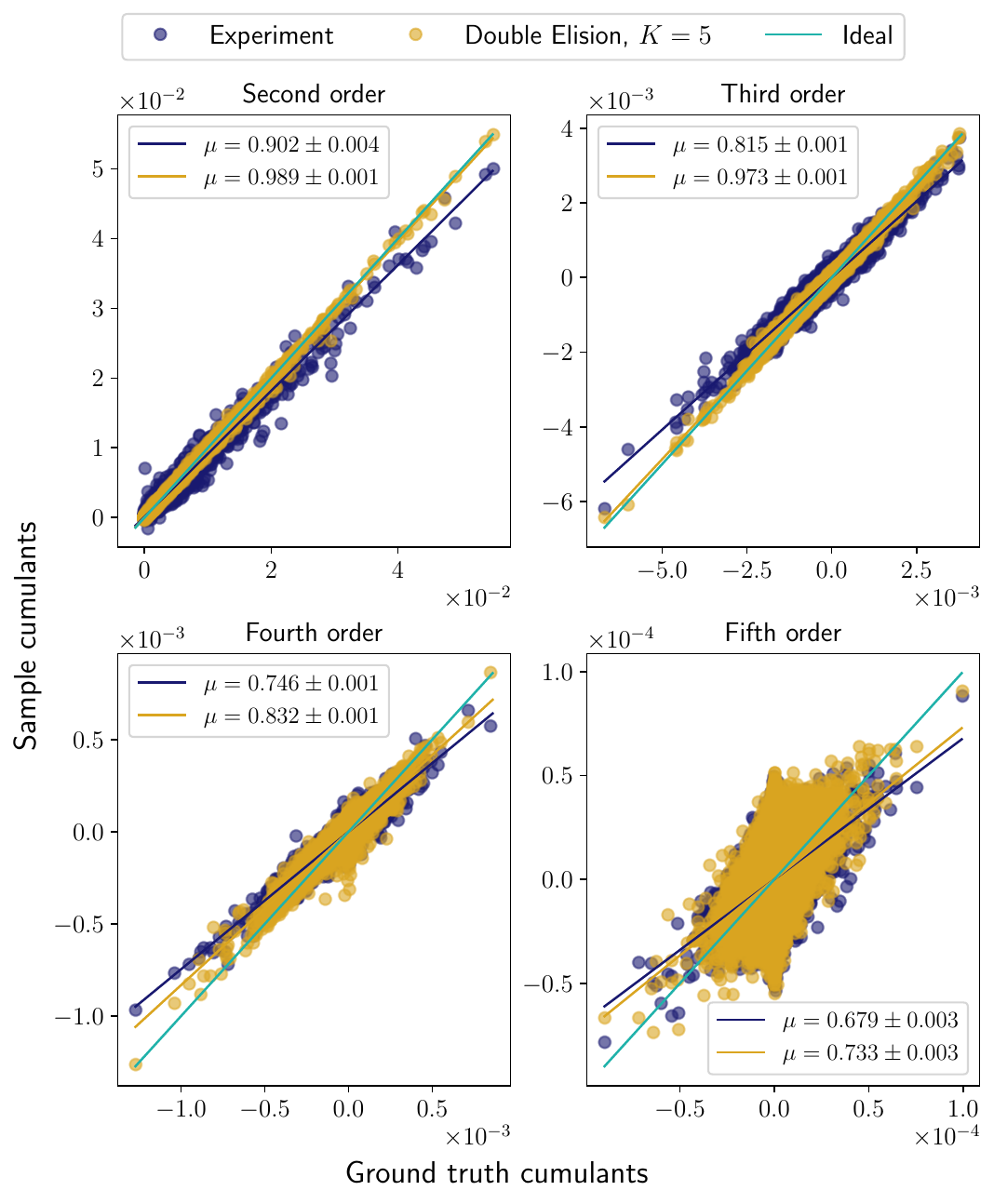}
         \end{subfigure}\\
         (a)
    \end{tabular}\qquad
    \begin{tabular}{@{}c@{}}
         \begin{subfigure}[c]{0.45\linewidth}
            \includegraphics[width=\textwidth]{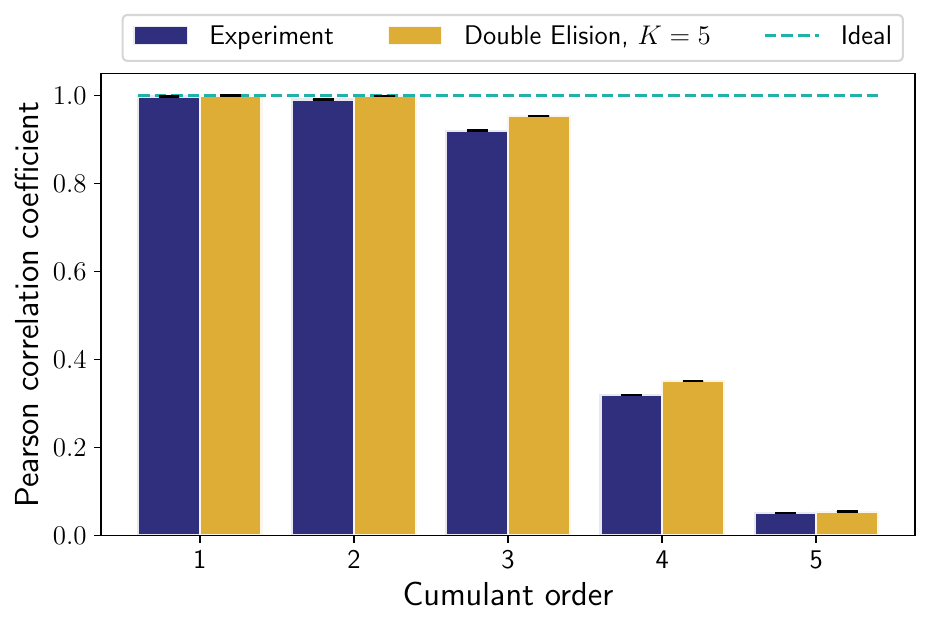}
         \end{subfigure}\\
         (b)\\
         \begin{subfigure}[c]{0.45\linewidth}
            \includegraphics[width=\textwidth]{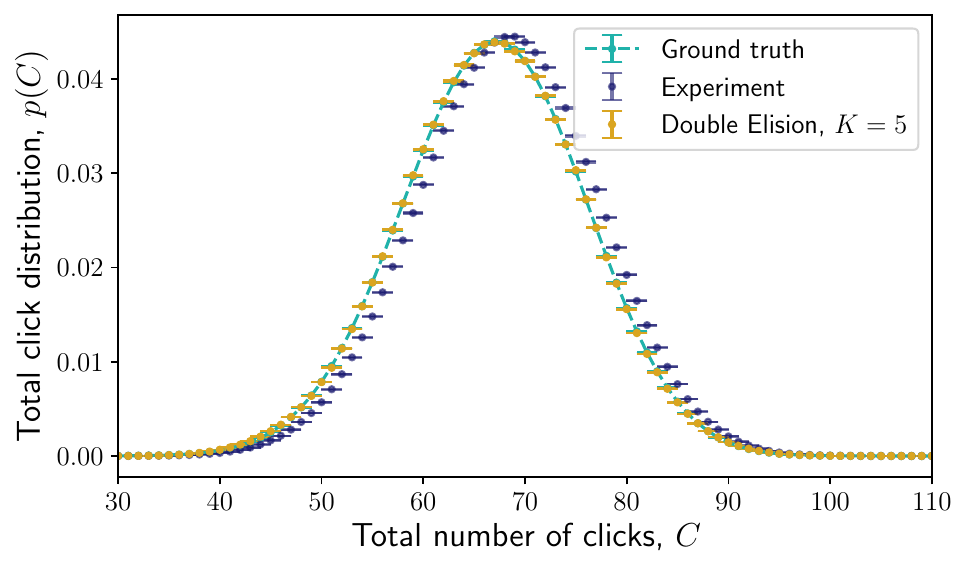}
         \end{subfigure}\\
         (c)\\
    \end{tabular}
    \caption{\textit{Validation tests for the J2-P65-5 experiment.} (a) Comparison between the estimated click cumulants (orders 2 to 5), from experimental (dark blue circles) and double-elision samples with $K=5$ (yellow circles), and those predicted by the ground truth of the experiment (light green line). The vertical axis corresponds to cumulants obtained from samples, either experimental or double-elision, while the horizontal axis corresponds to the ground truth (i.e., theoretical) cumulants. The dark blue and yellow lines represent the linear fit between sample cumulants and those from the ground truth. The slope of each linear fit is shown in the legends. All cumulants were estimated using $10^7$ samples. (b) Pearson correlation coefficients between sample and ground truth cumulants. The light green dashed line indicates the ideal value of these coefficients, which is equal to 1. Error bars were obtained using 100 bootstrapping resamples. (c) Total click distribution as a function of the total number of clicks $C$, for both experimental (dark blue circles) and double-elision (yellow circles) samples. The ground truth distribution corresponds to the light green dashed line with circles, which was obtained using phase space techniques~\cite{drummond2022simulating}. Error bars were obtained through bootstrapping.} 
    \label{fig:validation_J25}
\end{figure*}

\subsection{Validation on low-brightness}\label{subsec:lowbri}

\begin{figure*}[!ht]
    \centering
    \begin{tabular}{@{}c@{}}
         \begin{subfigure}[c]{0.45\linewidth}
            \includegraphics[width=\textwidth]{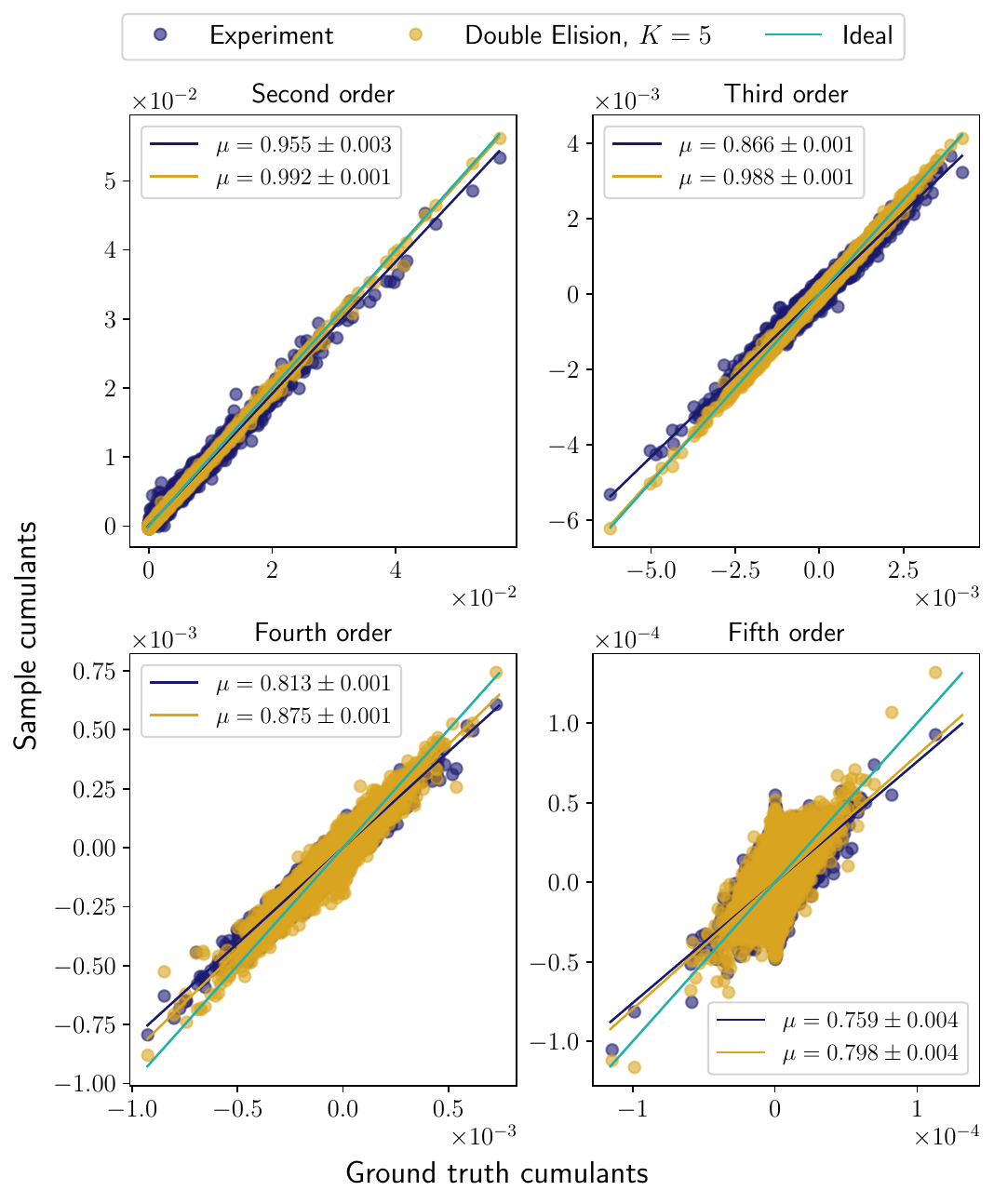}
         \end{subfigure}\\
         (a)
    \end{tabular}\qquad
    \begin{tabular}{@{}c@{}}
         \begin{subfigure}[c]{0.45\linewidth}
            \includegraphics[width=\textwidth]{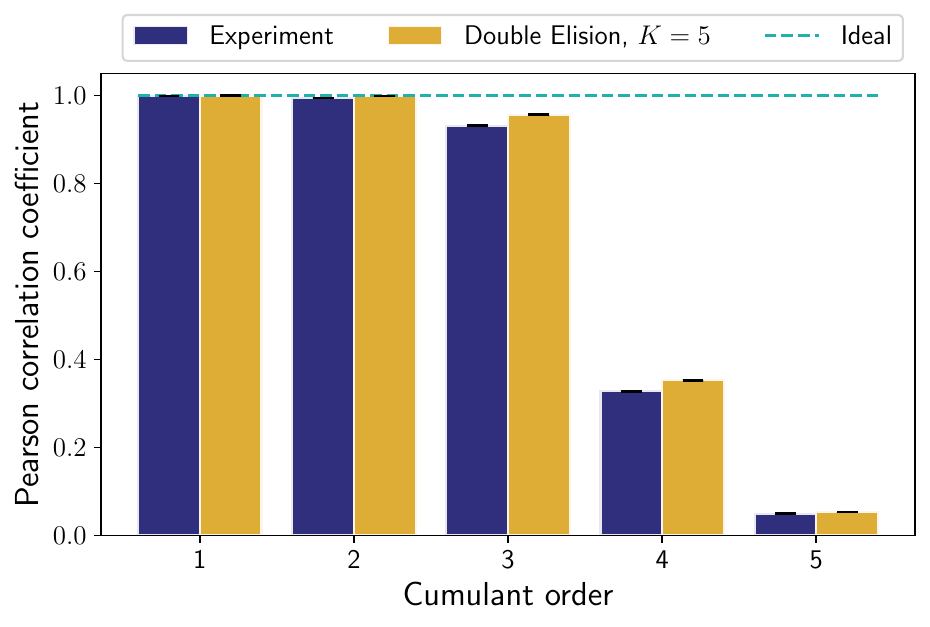}
         \end{subfigure}\\
         (b)\\
         \begin{subfigure}[c]{0.45\linewidth}
            \includegraphics[width=\textwidth]{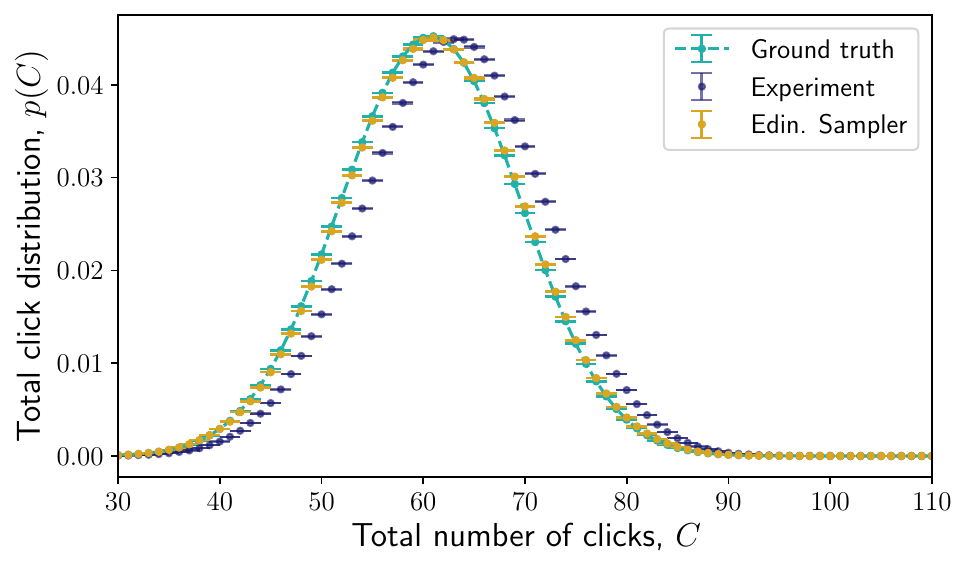}
         \end{subfigure}\\
         (c)\\
    \end{tabular}
    \caption{\textit{Validation tests for the J3-high experiment.} (a) Comparison between the estimated click cumulants (orders 2 to 5), from experimental (dark blue circles) and double-elision samples with $K=5$ (yellow circles), and those predicted by the ground truth of the experiment (light green line). The vertical axis corresponds to cumulants obtained from samples, either experimental or double-elision, while the horizontal axis corresponds to the ground truth (i.e., theoretical) cumulants. The dark blue and yellow lines represent the linear fit between sample cumulants and those from the ground truth. The slope of each linear fit is shown in the legends. All cumulants were estimated using $10^7$ samples. (b) Pearson correlation coefficients between sample and ground truth cumulants. The light green dashed line indicates the ideal value of these coefficients, which is equal to 1. Error bars were obtained using 100 bootstrapping resamples. (c) Total click distribution as a function of the total number of clicks $C$, for both experimental (dark blue circles) and double-elision (yellow circles) samples. The ground truth distribution corresponds to the light green dashed line with circles, and was obtained using phase space techniques~\cite{drummond2022simulating}. Error bars were obtained through bootstrapping. The estimation of the experimental and double-elision total click distributions was obtained using $10^7$ samples.}
    \label{fig:validation_J33}
\end{figure*}

Figure~\ref{fig:validation_J2_2} show our performance results
against a low-brightness configuration of the Jiuzhang 2.0 experiment, the configuration J2-P65-2 with an output photon-number density of $\nu = 0.093$. 
Due to the large number of modes in the system, verifying the proximity between the sampler's distribution and the ground truth of the experiment using the total variation distance, a standard statistical metric, cannot be done efficiently. Nevertheless, the low brightness of this configuration allows us to use XEB as a sample-efficient measure of similarity between the sampler's distribution and the ground truth as well as the calculation of click cumulants up to fifth order (see the Appendices Sec.~\ref{app:cumulants_benchmark} and Sec.~\ref{app:xeb} for description of the XEB score and click cumulants),
which have been widely-used verification techniques for recent implementations of GBS and the recent classical emulation in Ref.~\cite{oh2024classical}. 

The scatter plots in Figure~\ref{fig:validation_J2_2}(a) show the comparison between the estimated cumulants (orders 2 to 5) from both experimental and double-elision samples, and those predicted by the ground truth of the experiment. We estimated the cumulants using $10^7$ experimental and double-elision samples. The linear fits between sample and theoretical cumulants show that the emulator cumulants are better correlated with the theoretical values than those estimated from the experimental samples. Indeed, the slopes $\mu$ of the linear fits between double-elision and ground truth cumulants are closer to the ideal value of $\mu=1$ than those of the experimental cumulants. This is further confirmed by the Pearson correlation coefficients $r$ depicted in Figure~\ref{fig:validation_J2_2}(b): our sampler's correlation coefficients are higher and closer to the ideal value of $r=1$. 

In Figure~\ref{fig:validation_J2_2}(c) we show the XEB score as a function of the total number of clicks, $C$, for both experimental and double-elision samples. For reference, we also show the XEB scores for a squashed states model, which has previously been considered as a possible explanation for the Jiuzhang 2.0 experiment~\cite{martinez2023classical}. Each value of the XEB score was estimated using 4000 samples. We computed the XEB scores in the range $5\leq C\leq 20$, which, according to the inset of Figure~\ref{fig:validation_J2_2}(c), represents the most significant part of the total click distribution of J2-P65-2. As can be seen, the double-elision algorithm closely reproduces the XEB scores and for some values of $C$ outperforms the experimental samples scores.

\subsection{High-brightness results for Jiuzhang 2.0}

We now move on to the validation of our sampler for a Jiuzhang 2.0 configuration with high brightness, namely the J2-P65-5 configuration with $\nu = 0.975$. For this setup, the most relevant part of the total click distribution lies at $C>40$, making the XEB test too costly to carry out on the statistically relevant click counts. Similarly as in Ref.~\cite{oh2024classical}, we will rely solely on the comparison of click cumulants and the comparison of the total click distribution to investigate the performance of our sampler.

In Figure~\ref{fig:validation_J25}(a) we show the comparison between sample and ground truth cumulants (orders 2 to 5), as well as the corresponding linear fits. We computed the cumulants using $10^7$ experimental and double-elision samples. Again, we can readily see that, to all orders, our emulator's cumulants are better correlated with the ground truth than those estimated from the experimental samples. This is also confirmed by the Pearson correlation coefficients shown in Figure~\ref{fig:validation_J25}(b). Finally, Figure~\ref{fig:validation_J25}(c), shows a comparison of the total click distribution as a function of $C$. We can see that the distribution of the double-elision algorithm neatly overlaps with the ground truth distribution, while that of the experimental outcomes is slightly shifted.

\subsection{High-brightness results for Jiuzhang 3.0}

We also validate the performance of our sampler against the ``high power'' configuration of the Jiuzhang 3.0 experiment, i.e. J3-high, with a photon-number density of $\nu=0.81$. In this experiment, the input light propagates through a lossy interferometer with 144 spatial modes and is then time-domain multiplexed into 8 bins for readout using threshold detectors, in order to implement a pseudo-photon-number-resolution. This implies that the actual Gaussian state before measurement has $M=1152$ modes rather than $M=144$. Nevertheless, following a similar strategy as the one proposed in Ref.~\cite{oh2024classical}, we will verify our sampler against a ``coarse-grained'' configuration with $M=144$. As in the case of J2-P65-5, the computation of the XEB scores becomes too costly in the relevant range of the total number of clicks, $C>40$. Thus, we will focus on the comparison of click cumulants and distribution of the total number of clicks. Figure~\ref{fig:validation_J33} shows that the performance of our sampler for these coarse-grained tests is as good as or better than the experiment.

\section{Cost estimation and scaling}\label{sec:cost_and_scaling}

In what follows we present results that demonstrate the scaling of the different computational tasks of our procedure and its capability for parallelization. We used a CPU for this analysis, as we could explicitly run the tools with various numbers of cores. For a general use, a GPU implementation, as used in Sec. \ref{sec:comparison}, is recommended for a greater sampling speed.

\begin{figure*}[!t]
    \centering
    \begin{tabular}{@{}c@{}}
         \begin{subfigure}[c]{0.45\linewidth}
            \includegraphics[width=\textwidth]{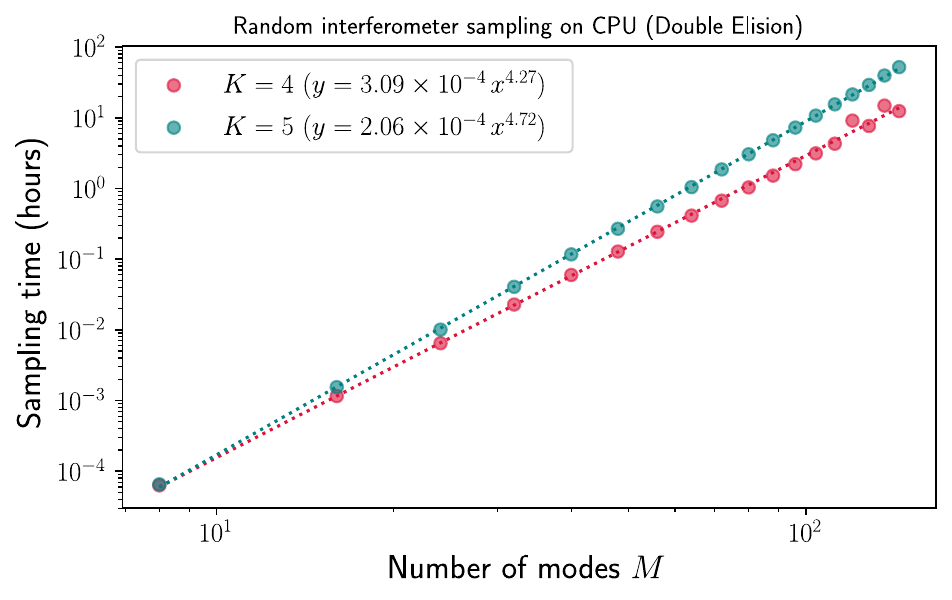}
         \end{subfigure}\\
         (a)\\
         \begin{subfigure}[c]{0.45\linewidth}
            \includegraphics[width=\textwidth]{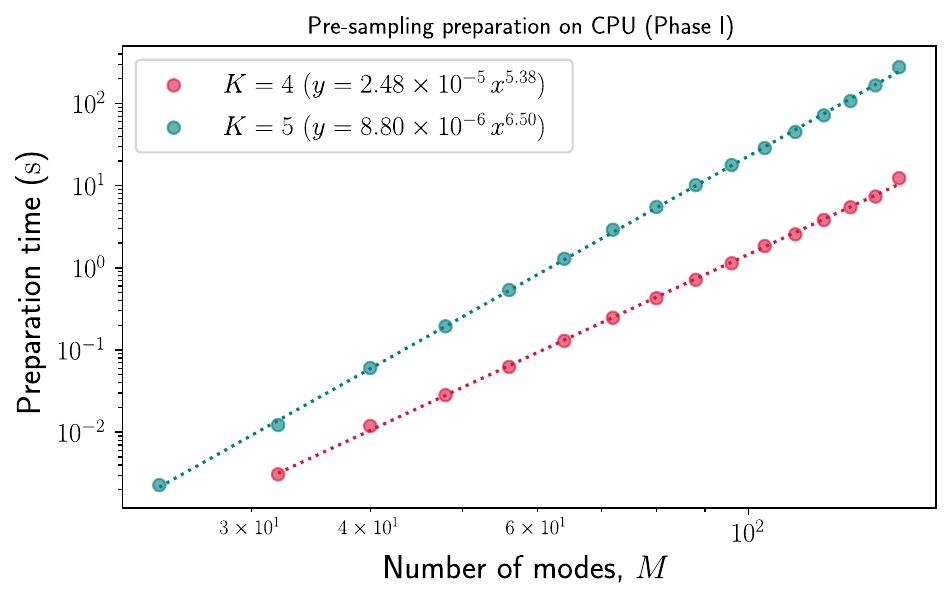}
         \end{subfigure}\\
         (c)\\
    \end{tabular}\qquad
    \begin{tabular}{@{}c@{}}
         \begin{subfigure}[c]{0.45\linewidth}
            \includegraphics[width=\textwidth]{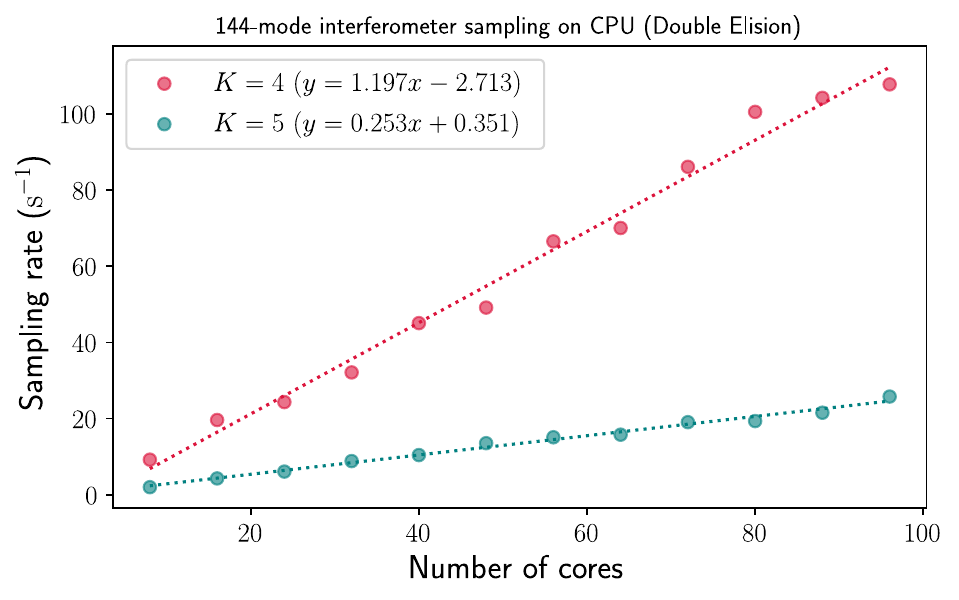}
         \end{subfigure}\\
         (b)\\
         \begin{subfigure}[c]{0.45\linewidth}
            \includegraphics[width=\textwidth]{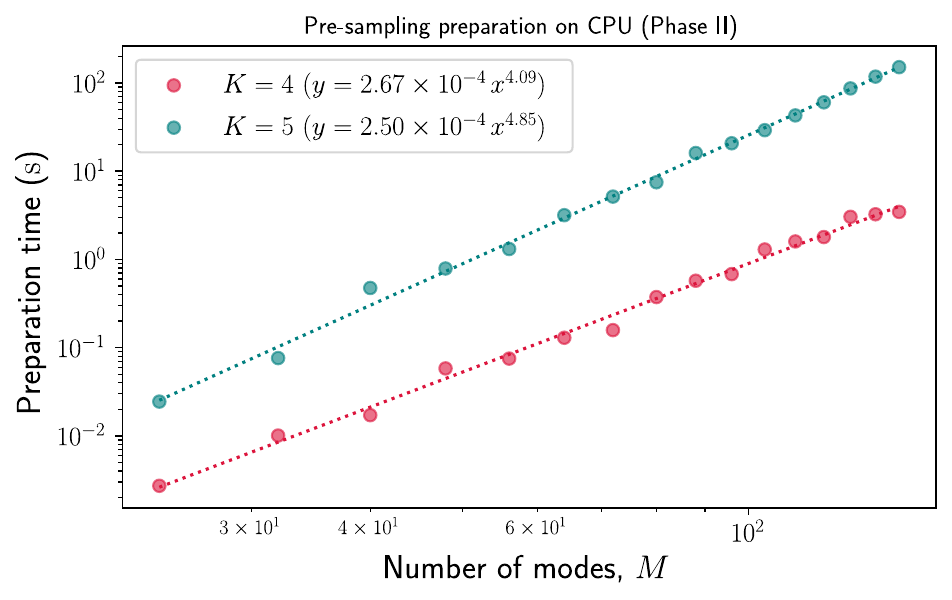}
         \end{subfigure}\\
         (d)\\
    \end{tabular}
    \caption{\textit{Double-elision algorithm cost estimation}. (a) Sampling time (in hours) as a function of the number of modes, $M$, for $K=4$ (red circles) and $K=5$ (teal circles). Each point indicates the time it takes to generate $10^7$ samples. (b) Sampling rate (in $\mathrm{s}^{-1}$) as a function of the number of cores for $K=4$ (red circles) and $K=5$ (teal circles). (c) Time of the Phase I, i.e. data-structure pre-processing, (in $\mathrm{s}$) as a function of the number of modes, $M$, for $K=4$ (red circles) and $K=5$ (teal circles). (d) Time of the Phase II pre-sampling preparation (in $\mathrm{s}$) as a function of the number of modes, $M$, for $K=4$ (red circles) and $K=5$ (teal circles). In Figs. (a) to (d), the dotted lines correspond to the linear or power law fits shown in the legends. In figures (a), (c) and (d), both the horizontal and vertical axes are in logarithmic scale. All data where obtained using randomly generated interferometers, except for (b) where we used the 144-mode interferometer of the J3-high configuration.}
    \label{fig:cost_estimation}
\end{figure*}

As discussed in Sec.~\ref{sec:new emulator}, our sampling algorithm should scale as $O(M^K)$ in time and space, where $M$ is the number of modes in the Gaussian state and $K$ is the fixed truncation order. Figure~\ref{fig:cost_estimation}(a), using a log-log scale, shows the sampling time of the algorithm as a function of the number of modes for truncation orders $K=4$ and $K=5$, from which we infer a good agreement with a polynomial scaling $O(M^K)$. The power is slightly higher for $K=4$ and lower for $K=5$, which is potentially an implementation artifact. Figure~\ref{fig:cost_estimation}(b) shows the sample generation rate as a function of the number of cores used. The clear linear relation for both $K=4$ and $K=5$, albeit with different slopes, demonstrates the embarrassingly parallelizable nature of the algorithm.

The pre-processing stage discussed in Sec.~\ref{subsec:Training} pre-computes the correlators and cumulants of a given ground truth, which are later used during the generation of samples. This procedure consists of two phases. The first phase is a \textit{data-structure pre-processing} task that only needs to be run once for each choice of $M$ and $K$, i.e. it does not change as we change the GBS circuit. It relates to Eq.~(\ref{eq:cumulant_corr_main}) and consists of obtaining all the sets of indices of the form $S=\{i_1,\dots,i_d\}\subseteq [M]$, with $d$ up to $K$, as well as on the computation of all necessary ``partition patterns'', which are used to identify any partition of a given $S$ in Eq.~(\ref{eq:cumulant_corr_main}). Each pattern corresponds to a partition $\pi$ of the set $\{1, \dots, d\}$, and a constant $w(\pi)=(-1)^{|\pi|-1}(|\pi|-1)!$ that determines the contribution of that partition to the estimation of the cumulants in the next phase of pre-processing. We remark that in the case of the statistical estimation of cumulants from sampled data, used in our click cumulant analysis, the weights $w(\pi)$ have an extra correction factor depending on the number of samples~\cite{smith2020tutorial} (see Sec. \ref{app:cumulants_benchmark}).

The second pre-processing stage consists of \textit{computing all the cumulants} of a given GBS up to order $K$ using Eq.~\eqref{eq:corre_cov_main}. Each block $b\in \pi$ is mapped to a correlator $c(\{i_k \text{ with } k\in b\})$. Then, these correlators are multiplied together according to $\pi$, weighted and summed in order to obtain the necessary cumulants following Eq.~\eqref{eq:cumulant_corr_main}. Figures~\ref{fig:cost_estimation}(c) and~\ref{fig:cost_estimation}(d) show the time complexity of Phase I and Phase II, respectively. Though Phase II approximately follows the expected time complexity of $O(M^K)$, Phase I scales more like $O(M^{K+1})$ or higher, which potentially hints at the need for an improvement in its implementation.

The space complexity of the pre-processing and sampling algorithm are more straightforward to analyse. The pre-sampling stage requires just enough memory to compute and store the cumulants subsequently used by the sampling algorithm. Although there are $\binom{M}{K}\approx O(M^K)$ cumulants in total, our implementation for computing the cumulants from the correlators also requires the use of an array of auxiliary indices of size $O(KM^K)$. For sampling, there are two dominant sources of memory cost: the pre-computed array of $O(M^K)$ cumulants and the dynamically-updated table of $O(M^3)$ approximated marginals during the generation of a sample.  For the 144-mode algorithm at $K=5$, the array of cumulants requires 2 GB, the auxiliary indices require 2.5 GB, and the table of marginals while sampling requires 2 MB per thread.

\section{Discussion}
\label{sec:Discussion}

We have proposed a GBS emulator based on an expansion in terms of cumulants up to a chosen truncation order. Its implementation is embarrassingly parallelizable and benefits from a polynomial scaling in the system size, with an exponent depending on the truncation order. We believe its characteristics make it a solid contender against future generations of GBS hardware. We also believe the algorithm is easily generalisable to photo-counting and single-photon source scenarios.

Our work shows that a strong requirement for hardness of GBS is the presence of non-trivial high-order cumulants, not just in the original noiseless distribution, but more importantly in the experimental implementation affected by losses and other imperfections. Therefore, hardware design should be oriented at carefully balancing the increase of modes while preserving or increasing the high-order correlations, as low order emulators are not difficult to implement and scale, as we have shown in this work. 

Losses and imperfections are not directly exploited in our algorithm's design beyond motivating the truncation of the cumulant expansion. Therefore, our algorithm would be able to simulate noiseless quantum circuits that have negligible cumulants beyond a constant order $K$, something that would not be possible with other emulators of GBS.

High-brightness sources impose surmountable challenges to our emulator, as they potentially increase the cumulant expansion truncation order and also require better marginal approximations due to the larger cumulant values. One should not immediately infer from this that increasing the brightness is a good strategy to preserve the hardness of simulation. We believe that adapting our algorithms to a decomposition of highly noisy states in terms of displaced pure states of low brightness, as done in Ref.~\cite{oh2024classical}, will support our intuition. 

To scale the number of modes into the thousands in order to simulate the very recent Jiuzhang 4.0 experiment \cite{liu2025}, we may need a new implementation that not only makes use of multiple nodes of an HPC cluster, but also accurately handles large sums of extremely small probabilities. At the same time, future algorithmic improvements, such as the decomposition of a highly noisy state in terms of displaced pure state, could downscale the costs. One may also expect a slightly more efficient computation of the Fourier coefficients, as was already achieved for Torontonians~\cite{quesada2018gaussian,kaposi2021polynomial}, or a direct computation of the cumulants avoiding the intermediate calculation of joint moments as is known for photon-number cumulants~\cite{cardin2024photon}.

Our algorithmic framework based on a cumulant expansion and approximated marginals is valid for any binary output probability distribution, rendering it applicable beyond the emulation of GBS. For example, it could be adapted to emulate qubit-based quantum circuits, such as IQC and QAOA.
To be applicable, the noiseless or noisy circuit needs to satisfy that: (i) its cumulants of order higher than a constant $K$ are negligible; (ii) its cumulants can be efficiently computed up to order $K$. For example, one could envision using our tools to de-quantize ``train classically deploy quantum'' proposals.
In cases where assumption (i) above holds but not (ii), one could still design classical surrogate models of quantum circuits. These are families of quantum-enhanced classical algorithms where samples from a quantum device are used to train a classical model (estimate the cumulants) that learns to replicate the behaviour of the quantum device using purely classical resources (our sampling algorithm).

\section{Methods}
\subsection{Marginals approximations}

As a pedagogical example, we present a simplified ``single-elision'' variant of the sampling algorithm at $K=3$ that has a reduced time complexity and space complexity of $O(M^3)$. The double-elision algorithm used to obtain the results in subsection \ref{subsec:lowbri} is presented in full detail in Appendix \ref{App:marginals}.

The algorithm can be understood as a dynamical program, where we keep track of three tables. The first table stores the auxiliary partial marginals $p^{(+)}(x_a,\ldots,x_b)$, containing all variables between $x_a$ and $x_b$, where $a>b$ and $a,b\in[n-1]$. The second table stores the auxiliary ``elided'' marginals $p^{(1)}\pars{x_a,\ldots,\widetilde{x}_e,\ldots,x_1}$, containing all variables between $x_1$ and $x_a$ except for the variable $x_e$ (we use $\widetilde{x}_e$ to denote that the variable is missing in the marginal), where similarly $a\in[n-1]$. Finally, the third table holds all of the sampling probabilities $p(x_a,\ldots,x_1)$, where again $a\in[n-1]$. With each generated bit, these tables are updated using a non-trivial combination of previously filled values and the pre-computed cumulants. 

Following Eq.~(\ref{eq:samplecumulant}), the joint probability $p(x_n,\ldots,x_1)$ of the first $n$ bits can be expanded up to order $K=3$,
{\small
\begin{align} \label{eq:methprob}
    &p(x_n,\ldots,x_1)  \\
    &=\frac{1}{2} \pars{1+\gamma_{\braces{n}}}\, p(x_{n-1},\ldots,x_1) \nonumber\\ 
    &+ \frac{1}{4} \sum_{i=0}^{n-1} \gamma_{\braces{i,n}}\,p^{(1)}\pars{x_{n-1},\ldots,\widetilde{x}_i,\ldots,x_1} \nonumber\\ 
    &+ \frac{1}{8} \sum_{\substack{i,j=1\\ i>j}}^{n-1} \gamma_{\braces{i,j,n}}\,
    p^{(+)}(x_{n-1},..,x_{i+1})\,p^{(1)}\pars{x_{i-1},..,\widetilde{x}_j,..,x_1}, \nonumber 
\end{align}
}
where each $\gamma_S$ is the product of the order-$|S|$ cumulant $\kappa\pars{S}$ and the associated parity function $\chi_S(\bar{x})$, and we only include the terms in the second sum for which $i>j$.

The second-order term in Eq.(\ref{eq:methprob}) has marginals with a single missing variable, which we approximate via the auxiliary marginal distribution 
{\small
\begin{align} \label{eq:b}
&p^{(1)}(x_n,\ldots,\widetilde{x}_{e},\ldots,x_1)\\
& = \frac{1}{2} \pars{1+\gamma_{\braces{n}}}\, p^{(1)}(x_{n-1},\ldots,\widetilde{x}_e,\ldots,x_1) \nonumber\\ 
    &+ \frac{1}{4} \sum_{\substack{i=1\\i\neq e}}^{n-1}
    \gamma_{\braces{i,n}}p^{(+)}\pars{x_{n-1},..,x_{b+1}}\, 
    p^{(1)}\pars{x_{b-1},..,\widetilde{x}_a,..,x_1}, \nonumber
\end{align}
}
where $a=\text{min}(i,e)$, $b=\text{max}(i,e)$ and we exclude the term in the sum for which $i=e$.
The third-order term in Eq.(\ref{eq:methprob}) does not contain a double-elided auxiliary marginal $p^{(2)}(x_{n-1},\ldots,\widetilde{x}_{b},\ldots,\widetilde{x}_{a},\ldots,x_1)$, as computing these would lead to a scaling in the time complexity beyond $O(M^3)$.
A similar case can be made for the second-order term in Eq.(\ref{eq:b}). 
We therefore use a cheaper approximation of those marginal terms via
product of two independent terms, one of which is the auxiliary partial marginal
{\small
\begin{align} \label{eq:a}
&p^{(+)}(x_n,\ldots,x_\ell) \\
&= \frac{1}{2} \pars{1+\gamma_{\braces{n}}}\, p^{(+)}(x_{n-1},\ldots,x_\ell) \nonumber\\ 
&+ \frac{1}{4} \sum_{i=\ell}^{n-1} \gamma_{\braces{i,n}}\,p^{(+)}\pars{x_{n-1},\ldots,x_{i+1}}\,p^{(+)}\pars{x_{i-1},\ldots,x_\ell} \nonumber.
\end{align}
}
The $n$th step of computing $p(x_n,\ldots,x_1)$ from previously estimated values involves a sum of 
$O(n^2)$ terms. For a bitstring of size $M$, the time complexity is $O(M^3)$.
A similar argument shows that computing all $p^{(+)}(x_n,\ldots,x_\ell)$ and $p^{(1)}(x_n,\ldots,\widetilde{x}_{e},\ldots,x_1)$ values has the same total time complexity of 
$O(M^3)$. These auxiliary marginals add a limited cost of storing $O(M^2)$ additional values related to the auxiliary marginal tables for each concurrent thread alongside the global $O(M^3)$ requirement of storing the cumulants up to order $K=3$.

\subsection{Implementation}
 For the statistical comparison to each experiment, we generated 10 million samples in 115 minutes using 8 Nvidia H100 SXM 80 GB GPUs on one cluster node. Each GPU used 2 GB of memory to store the list of cumulants up to fifth order, and the rest of the available memory was split into units of 2 MB to hold each launched thread's tables of marginals (each thread was responsible for generating a single sample). Each kernel run generated a batch of samples with a size determined by the available memory, which in this case was approximately 38000. In order to generate the complete list of samples, the kernel was launched multiple times across each GPU.

The benchmarks were obtained using a single AMD EPYC 9654 CPU of the Rorqual compute cluster at the \'Ecole de Technologie Sup\'erieure (ETS) at Montréal. 
For the cost estimation and scaling tests, we used a CPU implementation multithreaded with OpenMP. The threads read from a single shared list of cumulants, again requiring 2 GB of memory, and each used a private 2 MB cache to store the tables of marginals required for the current sample. Unlike the GPU implementation, each thread is responsible for iteratively generating multiple samples rather than just one, and the memory storing the tables of marginals can easily be continually reused.

\section{Acknowledgements}
N.Q. and J.M.-C. acknowledge the Alliance International Catalyst Quantum Grant (ALLRP) and the Discovery Grants program from NSERC of Canada; their research was enabled in part by support provided by the Digital Research Alliance of Canada and the Institut transdisciplinaire d’information quantique (INTRIQ), a Strategic Cluster funded by Fonds de recherche du Québec – secteur Nature et technologies.
The PhD Scholarship of T.D. is supported by Xanadu Quantum Technologies Inc. R.G.-P. was supported by the EPSRC-funded project Benchmarking Quantum Advantage.  
 R.G.-P. and T.D. thank Antonio Vergari, Lorenzo Loconte and Adrián Javaloy for fruitful discussions. We thank Zhenyan Zhao for early work 
 during his undergraduate project in 2022-23 under the supervision of R.G.-P. at the University of Edinburgh.

\bibliography{GBSemulator}

\onecolumngrid

\newpage

\appendix
\section*{Appendices}

\begin{itemize}
    \item[\ref{app:probs_fou}.] Probabilities, Fourier coefficients and cumulants.\hfill\pageref{app:probs_fou}
    \begin{itemize}
        \item[\ref{app:connect_to_spin}.] Connection to spin systems.\hfill\pageref{app:connect_to_spin}
        \item[\ref{app:mcc}.] Joint moments, cumulants and correlators.\hfill\pageref{app:mcc}
    \end{itemize}
    \item[\ref{app:probs_cumul}.] Probabilities expansion in terms of cumulants.\hfill\pageref{app:probs_cumul}
    \item[\ref{App:marginals}.] Approximating marginals efficiently in a chain rule.\hfill\pageref{App:marginals}
    \begin{itemize}
        \item[\ref{app:double_elision}.] Double-Elision Method $K=5$.\hfill\pageref{app:double_elision}
    \end{itemize}
    \item[\ref{app:corrs}.] Correlators and cumulants of GBS with binary outcomes.\hfill\pageref{app:corrs}
    \begin{itemize}
        \item[\ref{app:corr_gbs}.] Correlators of GBS with binary outcomes.\hfill\pageref{app:corr_gbs}
        \item[\ref{app:cumul_exp}.] cumulant expression.\hfill\pageref{app:cumul_exp}
    \end{itemize}
    \item[\ref{app:comparison}.] Comparison to other implementations and experiment.\hfill\pageref{app:comparison}
    \begin{itemize}
        \item[\ref{app:time_complexity}.] Time complexity.\hfill\pageref{app:time_complexity}
        \item[\ref{app:space_complexity}.] Space complexity.\hfill\pageref{app:space_complexity}
        \item[\ref{app:parallel_scalability}.] Parallel scalability.\hfill\pageref{app:parallel_scalability}
    \end{itemize}
    \item[\ref{app:benchmark}.] Benchmarking theory and implementation.\hfill\pageref{app:benchmark}
    \begin{itemize}
        \item[\ref{app:cumulants_benchmark}.]Click cumulants, Pearson and Spearman coefficients.\hfill\pageref{app:cumulants_benchmark}
        \item[\ref{app:total_clicks}.]Total click distribution.\hfill\pageref{app:total_clicks}
        \item[\ref{app:xeb}.] Cross-entropy benchmarking (XEB).\hfill\pageref{app:xeb}
    \end{itemize}
\end{itemize}

\section{Probabilities, Fourier coefficients and cumulants \label{app:probs_fou}}

A probability distribution $p(\bar{x})$ over the sample space of $n$ binary variables, $\bar{x}=(x_1,x_2,\ldots,x_n)$, is a non-negative Boolean function, $p(\bar{x}):\{0,1\}^n\rightarrow \mathbb{R}$ with input \textit{bitstring} of length $n$ and output a real number encoding the corresponding occurrence probability. Basics of Boolean Fourier Analysis \cite{odonnell2021} shows that $p(\bar{x})$ belongs to a Hilbert space of dimension $2^n$ with inner product given by
\begin{equation}
    \langle p(\bar{x}), q(\bar{x})\rangle=\frac{1}{2^n}\sum_{\bar{x}\in\{0,1\}^n}p(\bar{x})q(\bar{x}).
    \label{eq:inner_product_fourier}
\end{equation}
The parity functions
\begin{equation}
    \chi_{\bar{s}}(\bar{x})=(-1)^{\bar{s}\cdot\bar{x}},
    \label{eq:parity_functions_fourier}
\end{equation}
indexed by the bitstring $\bar{s}\in\{0,1\}^n$, where $\bar{s}\cdot\bar{x}=\sum_{k=1}^nx_ks_k$, form a basis of the Hilbert space and satisfy the orthogonality condition
\begin{equation}
    \sum_{\bar{x}\in\{0,1\}^n}\chi_{\bar{s}}(\bar{x})\chi_{\bar{s}'}(\bar{x})=2^n\delta_{\bar{s},\bar{s}'}.
    \label{eq:orthogonality_relations_fourier}
\end{equation}
Using this basis, we can expand the probability distribution as
\begin{equation}
    p(\bar{x}) = \sum_{\bar{s}\in\{0,1\}^n}\hat{p}(\bar{s})\chi_{\bar{s}}(\bar{x}),
    \label{eq:fourier_expansion_prob}
\end{equation}
where the $\hat{p}(\bar{s})=\langle p(\bar{x}), \chi_{\bar{s}}(\bar{x})\rangle$ are known as \textit{Fourier coefficients}. The following inverse relation also holds
\begin{equation}
    \hat{p}(\bar{s}) = \frac{1}{2^n}\sum_{\bar{x}\in\{0,1\}^n}p(\bar{x})\chi_{\bar{s}}(\bar{x})
    \label{eq:inverse_fourier_expansion_prob}
\end{equation}
The quantity $|\bar{s}|=\sum_{i=1}^ns_i$ represents the \textit{Hamming weight} of the Fourier coefficient.

Note that every $\bar{s}$ is uniquely determined by the number of $1$'s it has (i.e. its Hamming weight) and by their positions within the bitstring. These positions can be parametrized by a subset $S\subseteq [n]=\{1,\dots,n\}$, while the Hamming weight is simply given by the cardinality of the subset, noted $|S|$. This allows us to associate every $\bar{s}\in\{0,1\}^n$ to a unique element of $\Omega([n])$, the power set of $[n]$. On this account, we can recast Eqs.~\eqref{eq:fourier_expansion_prob} and ~\eqref{eq:inverse_fourier_expansion_prob} as
\begin{equation}
    p(\bar{x}) = \sum_{S\subseteq [n]}\hat{p}(S)\chi_{S}(\bar{x})\;,\qquad \hat{p}(S) = \frac{1}{2^n}\sum_{\bar{x}\in\{0,1\}^n}p(\bar{x})\chi_{S}(\bar{x}),
    \label{eq:fourier_relations_subset}
\end{equation}
where we redefine the parity functions as
\begin{equation}
    \chi_{S}(\bar{x})=(-1)^{\sum_{k\in S}x_k} = \prod_{k \in S}(-1)^{x_k}.
    \label{eq:parity_functions_redefined}
\end{equation}

\subsection{Connection to spin systems \label{app:connect_to_spin}}
Inspecting the definition of the Fourier coefficients above, as well as that of the parity functions, we can recognize that the $\hat{p}(S)$ are, up to a $2^{-n}$ constant, equivalent to the definition of the \textit{spin correlators} of a system of $n$ spins defined by the functions $\{(-1)^{x_k}\}_{k\in S}$. The precise definition of these correlators reads
\begin{equation}
    c(S)= \mathbb{E}\left[\prod_{k \in S}(-1)^{x_k}\right]=\sum_{\bar{x}\in\{0,1\}^n}p(\bar{x})\prod_{k \in S}(-1)^{x_k}=2^n\hat{p}(S),
    \label{eq:correlators_def_bitstrings}
\end{equation}
where $\mathbb{E}[\cdot]$ stands for the expectation value with respect to the probability distribution $p(\bar{x})$.

\subsection{Joint moments, cumulants and correlators\label{app:mcc}}

Consider the random variables $\{x_k\}_{k\in S}$ with $S\subseteq [n]$. The joint moment of the $\{x_k\}_{k\in S}$ with respect to the distribution $p(\bar{x})\equiv p(x_1, \dots, x_n)$ is defined as 
\begin{equation}
    \mu(\{x_k\}_{k\in S})= \mathbb{E}\left[\prod_{k \in S}x_k\right]=\sum_{\bar{x}\in\{0,1\}^n}p(\bar{x})\prod_{k\in S} x_{k}.
    \label{eq:joint_moments_def}
\end{equation}
Since each of the $x_k$ can only take values in $\{0, 1\}$, we can readily see that the joint moments can be computed as the marginal probabilities $\mu(S)=p(x_k=1\text{ for }k\in S)$, i.e. 
the probability that all bits belonging to the set $S$ have value one. 

Associated to each joint moment, we can define the cumulant \cite{fisher1932derivation} (these quantities are also known in the literature as Ursell functions~\cite{ursell_1927}, truncated correlation functions~\cite{duneau1973decrease} or cluster functions~\cite{duneau1973decrease}) of the $\{x_k\}_{k\in S}$ as
\begin{equation}
    \tilde{\kappa}(\{x_k\}_{k\in S})= \sum_{\pi \in \mathcal{P}(S)}(-1)^{|\pi|-1}(|\pi|-1)!\prod_{b\in \pi}\mu(\{x_k\}_{k\in b}).
    \label{eq:joint_cumulants_def}
\end{equation}
In this expression, $\pi$ is a set partition of $S$, i.e. a collection of subsets of $S$ ($\pi=\{b_1,\dots,b_\ell\}$, $b_k\subseteq S$ for all $k$), usually referred to as ``blocks'', that are non-empty ($b_k\neq \emptyset$ for all $k$), mutually disjoint ($b_j\cap b_k= \emptyset$ for $j\neq k$), and whose union is equal to $S$ ($\bigcup_{b\in \pi}b = S$). The quantity $|\pi|$ indicates the number of blocks of the partition, while $\mathcal{P}(S)$ stands for the set of all partitions of $S$. Relation \eqref{eq:joint_cumulants_def} can be inverted to obtain the joint moments as a function of the cumulants:
\begin{equation}
    \mu(\{x_k\}_{k\in S})= \sum_{\pi \in \mathcal{P}(S)}\prod_{b\in \pi}\tilde{\kappa}(\{x_k\}_{k\in b}).
    \label{eq:moments_cumulants_rel}
\end{equation}

With these definitions in mind, we can readily see that the correlators $c(S)$ can be interpreted as the joint moments of the variables $\{(-1)^{x_k}\}_{k\in S}$, i.e. $c(S) =\mu(\{(-1)^{x_k}\}_{k\in S})$. We write the corresponding cumulants as $\kappa(S) \equiv \tilde{\kappa}(\{(-1)^{x_k}\}_{k\in S})$. Since each $x_k$ is a binary random variable, it holds that $(-1)^{x_k} = 1-2x_k$, which in turn leads to the relation 
\begin{equation}
    \kappa(S) =
    \begin{cases}
        1 - 2\mathbb{E}[x_k], & \text{for }|S| = 1,\\
        (-2)^{|S|} \tilde{\kappa}(\{x_k\}_{k\in S}), & \text{for }|S| > 1,
    \end{cases}
    \label{eq:cumulant_exponential}
\end{equation} 
where we have used the homogeneity and translation invariance properties of the cumulants. Following this result, we can compute the correlators $c(S)$ as
\begin{equation}
    c(S)= \sum_{\pi \in \mathcal{P}(S)}\prod_{b\in \pi}\kappa(b).
    \label{eq:correlators_cumulants_rel}
\end{equation} 

\section{Expansion of probabilities in terms of cumulants\label{app:probs_cumul}}

Consider the expression for the probability distribution, $p(\bar{x})=p(x_1, \dots, x_n)$, in terms of the correlators $c(S)$:
\begin{equation}
    p(x_1, \dots, x_n)=\frac{1}{2^n}\sum_{S\subseteq[n]}\chi_S(\bar{x})c(S).
    \label{eq:prob_dist_corr}
\end{equation}
We want to change the sum over $S\subseteq [n]$ for a sum over $S'\subseteq [n-1]$. To do this, notice that all the subsets of $[n]$ can be divided into two groups: those that contain the element $n$, and those that do not. The collection of subsets that do not contain the element $n$ corresponds precisely to the power set of $[n-1]$, while every subset of $[n]$ containing $n$ can be written as $S'\cup\{n\}$, with $S'\subseteq[n-1]$. With these ideas in mind, we can recast Eq.~\eqref{eq:prob_dist_corr} as
\begin{align}
    p(x_1,\dots,x_n)&=\frac{1}{2^{n}}\sum_{S'\subseteq[n-1]} \chi_{S'}(\bar{x})\,c(S') + \frac{1}{2^{n}}\sum_{S'\subseteq [n-1]}\chi_{S'\cup\{n\}}(\bar{x})c(S'\cup\{n\}).
    \label{eq:probs_corr_proof_1}
\end{align}
Note that $\chi_{S'}(\bar{x})=\chi_{S'}(\bar{x}')$, with $\bar{x}'=(x_1,\dots,x_{n-1})$, since it is not possible for $x_n$ to appear in the sum $\sum_{k\in S'}x_k$ that defines $\chi_{S'}(\bar{x})$. Moreover, we have that $\chi_{S'\cup\{n\}}(\bar{x})=(-1)^{x_n}\chi_{S'}(\bar{x}')$. Taking into account that $p(\bar{x}') = p(x_1,\dots,x_{n-1}) = 2^{-(n-1)}\sum_{S'\subseteq[n-1]} \chi_{S'}(\bar{x}')\,c(S')$, we can write the probability distribution as
\begin{align}
    p(x_1,\dots,x_n)&=\frac{1}{2}p(x_1,\dots,x_{n-1}) + \frac{(-1)^{x_n}}{2^{n}}\sum_{S'\subseteq [n-1]}\chi_{S'}(\bar{x}')c(S'\cup\{n\}).
    \label{eq:probs_corr_proof_2}
\end{align}

Let us focus now on the term $c(S'\cup\{n\})$. According to Eq.~\eqref{eq:correlators_cumulants_rel}, $c(S'\cup\{n\})$ can be computed as a sum of cumulants over all possible set partitions of $S'\cup\{n\}$. An arbitrary partition, $\pi$, of this set will have the form $\pi=\pi'\cup\{R \cup \{n\}\}$, where $R\subseteq S'$, and $\pi'$ is a partition of $S'\setminus R$. We can see this by noticing that any partition of $S'\cup\{n\}$ necessarily has a block that contains $n$. This block can contain other elements of $S'$, so it will generally have the form $\{R\cup\{n\}\}$ with $R\subseteq S'$. Since all the blocks in the partition must be mutually disjoint, all the blocks that are different from $\{R\cup\{n\}\}$ must not contain any of the elements in $R$. Thus they will form a collection of non-empty, mutually disjoint subsets of $S'\setminus R$ whose union is precisely $S'\setminus R$, that is, a partition of $S'\setminus R$. With this result in mind, we can see that 
\begin{equation}
    \sum_{\pi\in \mathcal{P}(S'\cup \{n\})}\equiv \sum_{R\subseteq S'}\sum_{\pi'\in \mathcal{P}(S'\setminus R)}.
    \label{eq:sum_over_partitions_decomposition}
\end{equation}

This allows us to write
\begin{align}
    \label{eq:correlators_expanded}
    c(S'\cup\{n\}) &=\sum_{R\subseteq S'}\sum_{\pi'\in \mathcal{P}(S'\setminus R)}\prod_{b\in \pi'\cup\{R \cup \{n\}\}}\kappa(b)\nonumber\\
    &=\sum_{R\subseteq S'}\kappa(R\cup\{n\})\sum_{\pi'\in \mathcal{P}(S'\setminus R)}\prod_{b\in \pi'}\kappa(b).
\end{align}
Since $\sum_{\pi'\in \mathcal{P}(S'\setminus R)}\prod_{b\in \pi'}\kappa(b) = c(S'\setminus R)$, we obtain the following expression for $c(S'\cup\{n\})$:
\begin{align}
    c(S'\cup\{n\}) =\sum_{R\subseteq S'}\kappa(R\cup\{n\})\,c(S'\setminus R).
    \label{eq:correlators_expanded_2}
\end{align}

Combining Eqs.~\eqref{eq:probs_corr_proof_2} and~\eqref{eq:correlators_expanded_2}, we may write
\begin{equation}
    p(x_1,\dots,x_n)=\frac{1}{2}p(x_1,\dots,x_{n-1}) + \frac{(-1)^{x_n}}{2^{n}}\sum_{S'\subseteq [n-1]}\chi_{S'}(\bar{x}')\sum_{R\subseteq S'}\kappa(R\cup\{n\})c(S'\setminus R).
    \label{eq:probs_corr_cumulants}
\end{equation}
We can interchange the sums over $S'\subseteq[n-1]$ and $R\subseteq S'$ by noticing that these conditions imply that $R\subseteq [n-1]$, so we can sum over $R$ instead of $S'$ if we take into account that the final contribution of each $R$ is obtained from all the $S'\subseteq [n-1]$ such that $R\subseteq S'$. In other words, we have that
\[\sum_{S'\subseteq[n-1]}\sum_{R\subseteq S'}\equiv \sum_{R\subseteq [n-1]}\sum_{\substack{S'\subseteq [n-1]\\\text{s.t. }R\subseteq S'}}.\]
This allows us to write 
\begin{equation}
    \sum_{S'\subseteq [n-1]}\chi_{S'}(\bar{x}')\sum_{R\subseteq S'}\kappa(R\cup\{n\})c(S'\setminus R)=\sum_{R\subseteq [n-1]}\kappa(R\cup\{n\})\sum_{\substack{S'\subseteq [n-1]\\\text{s.t. }R\subseteq S'}}\chi_{S'}(\bar{x}')c(S'\setminus R).
    \label{eq:subsum_subsets}
\end{equation}
Noticing that every subset $S'\subseteq[m-1]$ containing $R$ can be written as $S'=T\cup R$, with $T\subseteq [n-1]\setminus R$, we can write 
\begin{equation}
    \sum_{\substack{S'\subseteq [n-1]\\\text{s.t. }R\subseteq S'}}\chi_{S'}(\bar{x}')c(S'\setminus R)=\sum_{T\subseteq [n-1]\setminus R}\chi_{T\cup R}(\bar{x}')c(T).
    \label{eq:changing_subsets}
\end{equation}

Eq.~\eqref{eq:changing_subsets} implies that Eq.~\eqref{eq:probs_corr_cumulants} will take the form
\begin{equation}
    p(x_1,\dots,x_n)=\frac{1}{2}p(x_1,\dots,x_{n-1}) + \frac{1}{2^{n}}\sum_{R\subseteq [n-1]}\chi_{R\cup\{n\}}(\bar{x}_{R\cup\{n\}})\kappa(R\cup\{n\})\sum_{T\subseteq [n-1]\setminus R}\chi_{T}(\bar{x}_T)c(T),
    \label{eq:probs_corr_cumulants_2}
\end{equation}
where $\bar{x}_A = (x_k \text{ for } k\in A)$, and we took into account that $(-1)^{x_n}\chi_{S'}(\bar{x}') = \chi_{R\cup\{n\}}(\bar{x}_{R\cup\{n\}})\chi_{T}(\bar{x}_{T})$. Introducing the marginal probability $p(\bar{x}_A)$, satisfying the relation
\begin{equation} 
    p(\bar{x}_A)=\frac{1}{2^{|A|}}\sum_{S\subseteq A} \chi_S(\bar{x}_A)c(S),
    \label{eq:probs_corr_proof_marginal}
\end{equation}
we obtain
\begin{equation}
    \label{eq:probs_corr_cumulants_3}
    p(x_1,\dots,x_n)=\frac{1}{2}p(x_1,\dots,x_{n-1}) + \sum_{R\subseteq [n-1]}\frac{1}{2^{|R|+1}}\chi_{R\cup\{n\}}(\bar{x}_{R\cup\{n\}})\kappa(R\cup\{n\})\,p(\bar{x}_{[n-1]\setminus R}).
\end{equation}

Using $p(x_1,\dots,x_{n-1})=p(x_1,\dots,x_{n-1},x_n)+p(x_1,\dots,x_{n-1},\bar{x_n})$ one can write the 
bias $\Delta_{x_n}=p(x_1,\dots,x_{n-1},x_n)-p(x_1,\dots,x_{n-1},\bar{x_n})$ as
\begin{equation}
    \label{eq:delta_bias}
    \Delta_{x_n}=\sum_{R\subseteq [n-1]}\frac{1}{2^{|R|}}\chi_{R\cup\{n\}}(\bar{x}_{R\cup\{n\}})\kappa(R\cup\{n\})\,p(\bar{x}_{[n-1]\setminus R}).
\end{equation}

This relation allows us to recursively compute the occurrence probability of a given bitstring using the cumulants of the distribution and previously computed marginals.

\section{Approximating marginals efficiently with a chain rule}\label{App:marginals}

Before going into the details below, we need to clarify that in this section we will invert the order of variables in $p(x_1,x_2,\dots,x_n)$ and instead use $p(x_n,x_{n-1},\dots,x_1)$, which is more natural for chain rule expressions related to conditional probabilities of the form $p(x_n|x_{n-1},\dots,x_1)$. In addition,
we will first need a few auxiliary definitions:
\begin{itemize}
    \item We will define the quantity $\gamma_A=\kappa(A)\chi(A)$, where
$\chi(A)=(-1)^{\sum_{i\in A}x_i}$ is the parity function over the set of variables $A$
and $\kappa(A)$ is its corresponding cumulant. 
\item We will need to use a compact notation for a particular form of sum over indices which are ordered alongside some other set of fixed indices which are not summed over. We will use
\begin{equation}
    \sum_{\substack{(i_1>i_2>\ldots>i_k)\in[m,n-1]\setminus\{e_1,e_2,\ldots,e_\ell\}\\ (a_1,a_2,\ldots,a_{k+\ell})={\rm ord}(i_1,i_2,\ldots,i_k,e_1,e_2,\ldots,e_\ell)}} 
\end{equation}
for a sum running over indices $i_1>i_2>\ldots>i_k$, ranging from $m$ to $n-1$ apart from the elided indices $e_1,e_2,\ldots,e_\ell$, where we relabel the summed and elided indices together as $a_1,a_2,\ldots,a_{k+\ell}$ by rearranging them in decreasing order. 

\item We will use the expression $\sumbinom{x}{d}$ as compact notation for
\begin{equation}
    \sumbinom{x}{d}=\sum_{i=0}^d\binom{x}{i}\approx \left(1+\frac{d}{x}\right)\frac{x^d}{d!}
\end{equation}
for the counting of cumulants up to order $d$. We will give details later on.
\end{itemize}

\subsection{Double-elision method $K=5$\label{app:double_elision}}

The expansion of the probability $p(x_n,\bar{x})$ up to fifth order ($K=5$) requires the use of three families of approximated marginals that are computed alongside $p(x_n,\ldots, x_1)$:

\begin{itemize}
    \item Marginals $p^{(+)}(x_{b},\ldots,x_{a})$ containing all variables from $x_a$ to $x_b$. We used the notation ``$(+)$'' to differentiate from the case where the variable with the lowest index is $x_{1}$, i.e. $p(x_n,\ldots,x_1)$. 
    \item Marginals with all variables between $x_{1}$ and $x_a$ except for one elision at $x_e$, i.e. $p^{(1)}\pars{x_{a},\ldots,\widetilde{x}_e,\ldots,x_1}$. We remind the reader that the tilde in $\widetilde{x}_e$ means that variable $x_e$ is not present in the marginal (tracing out the variable).
    \item Marginals with all variables between $x_{1}$ and $x_a$ except for two elisions at $x_d$ and $x_e$. i.e. $p^{(2)}\pars{x_{a},\ldots,\widetilde{x}_e,\ldots,\widetilde{x}_d,\ldots,x_1}$. 
\end{itemize}

The expansion of the probability $p(x_n,\ldots, x_1)$ up to fifth order ($K=5$) then reads:
\begin{align}
    \label{eq:probade}
    &p(x_n,\ldots, x_1) \nonumber\\& =\frac{1}{2} \pars{1+\gamma_{\braces{n}}}\, p(x_{n-1},\ldots,x_1) \nonumber\\ 
    &+\frac{1}{4} \sum_{i=1}^{n-1} \gamma_{\braces{i,n}}\,p^{(1)}\pars{x_{n-1},\ldots,\widetilde{x}_i,\ldots,x_1} \nonumber\\ 
    &+\frac{1}{8} \sum_{i=2}^{n-1} \sum_{j=1}^{i-1} \gamma_{\braces{i,j,n}}\pars{x_{n-1},\ldots,\widetilde{x}_i,\ldots,\widetilde{x}_j,\ldots,x_1} \nonumber\\ 
    &+\frac{1}{16} \sum_{i=3}^{n-1} \sum_{j=2}^{i-1} \sum_{k=1}^{j-1} \gamma_{\braces{i,j,k,n}}\,
    p^{(+)}(x_{n-1},\ldots,x_{i+1})\,p^{(2)}\pars{x_{i-1},\ldots,\widetilde{x}_j,\ldots,\widetilde{x}_k,\ldots,x_1}, \nonumber\\ 
    &+\frac{1}{32} \sum_{i=4}^{n-1}\sum_{j=3}^{i-1} \sum_{k=2}^{j-1} \sum_{\ell=1}^{k-1} \gamma_{\braces{i,j,k,\ell,n}}\,
    p^{(+)}(x_{n-1},\ldots,x_{i+1})\,p^{(+)}(x_{i-1},\ldots,x_{j+1})\,p^{(2)}\pars{x_{j-1},\ldots,\widetilde{x}_k,\ldots,\widetilde{x}_\ell,\ldots,x_1},
\end{align}
The approach is to replace the marginal $p(\bar{R})$, defined over the complement of the variables $R$, by an approximation. When the size of $R$ is one or two we can use the marginals with one or two elisions, respectively, i.e. $p^{(1)}$ and $p^{(2)}$. Beyond this, we simplify the marginal by splitting it into a product of independent terms, such that they completely cover the set of variables $R$. 

Our algorithm can be understood as a dynamical program, where we keep track of four tables, one for each type of approximated marginal $p^{(+)}(x_{b},\ldots,x_{a})$, $p^{(1)}\pars{x_{a},\ldots,\widetilde{x}_e,\ldots,x_1}$ and $p^{(2)}\pars{x_{a},\ldots,\widetilde{x}_e,\ldots,\widetilde{x}_d,\ldots,x_1}$, as well as the sampling probabilities $p(x_n,\bar{x})$. Each table is updated from a non-trivial combination of previously computed values.

At every step there is one term $p(x_n,\ldots,x_1)$ to compute consisting of a sum of size $\sumbinom{n-1}{4}\approx O(n^4)$. The largest sum update is of order $O(M^4)$, and
its total computing time $\sum_{i=1}^M\sumbinom{i-1}{4} \approx O(M^5)$ operations, requiring only a memory storage of $O(M)$. This does not finalize the cost analysis, as we need to incorporate the cost of the approximated marginals that we detail below and we finalize the section with a summary of the total costs.

As we will see below a key aspect of the design of the approximated marginals is to 
ensure the cost of filling the tables corresponding to the approximated marginals do not degrade the asymptotic scaling of $O(n^4)$ for the $n$th time step and a final time asymptotic of $O(M^5)$. 
This is achieved at the cost of storing $O(M^3)$ tables data during the dynamical program for each concurrent thread alongside the storage of the $\binom{M}{5}\approx O(M^5)$ global cumulants of order up to $K=5$.

\subsubsection{Marginal with non-fixed last variable}

The approximated marginals $p^{(+)}(x_{n},\ldots,x_{\ell})$ containing all variables between $x_{n}$ and $x_{\ell}$ being present is approximated by the following update rule:

\begin{align} \label{eq:probpartialade}
&p^{(+)}(x_n,\ldots,x_\ell)\nonumber\\ &= \frac{1}{2} \pars{1+\gamma_{\braces{n}}}\, p^{(+)}(x_{n-1},\ldots,x_\ell) \nonumber\\ 
&+ \frac{1}{4} \sum_{i=\ell}^{n-1} \gamma_{\braces{i,n}}\,p^{(+)}\pars{x_{n-1},\ldots,x_{i+1}}\,p^{(+)}\pars{x_{i-1},\ldots,x_\ell} \nonumber\\ 
&+ \frac{1}{8} \sum_{i=\ell+1}^{n-1} \sum_{j=\ell}^{i-1} \gamma_{\braces{i,j,n}}\,p^{(+)}\pars{x_{n-1},\ldots,x_{i+1}}\,p^{(+)}\pars{x_{i-1},\ldots,x_{j+1}}\,p^{(+)}\pars{x_{j-1},\ldots,x_\ell}
\end{align}

At every step of the chain rule ($n$th step), there are $n-1$ terms $p^{(+)}(x_n,\ldots,x_\ell)$ to compute, as $\ell\in[1,n-1]$:
\begin{itemize}
    \item Each term consists of a sum of size $\sumbinom{n-1}{2}\approx n^2/2$.
    \item A total of at most $5(n-1)\sumbinom{n-1}{2} \approx O(n^3)$ computing operations. 
\end{itemize}

In the process of generating a sample of size $M$ we will need:
\begin{itemize}
    \item A table of total size $\binom{M}{2}$ terms, resulting from having two free parameters ($n,\ell$) ranging from 1 to $M$.
    \item The longest update has a time complexity of $(M-1)\sumbinom{M-1}{2} \approx O(M^3)$.
    \item The total computing time is $\sum_{i=1}^M (i-1)\sumbinom{i-1}{2} \approx O(M^4)$ operations.
\end{itemize}

\subsubsection{Marginal with one elided variable}

The approximated marginals $p^{(1)}\pars{x_{n},\ldots,\widetilde{x}_e,\ldots,x_1}$
with all variables between $x_{n}$ and $x_{1}$ being present except for the elided variable
$x_e$ with index $e\in[1,n-1]$ can be computed from the following update rule:

\begin{align} \label{eq:probelidedade}
    &p^{(1)}(x_n,\ldots,\widetilde{x}_{e},\ldots,x_1)\nonumber\\ &= \frac{1}{2} \pars{1+\gamma_{\braces{n}}}\, p^{(1)}(x_{n-1},\ldots,\widetilde{x}_e,\ldots,x_1) \nonumber\\ 
    &+ \frac{1}{4} \sum_{\substack{i\in[1,n-1]\setminus\{e\}\\ (a,b)={\rm ord}(i,e)}} \gamma_{\braces{i,n}}\,p^{(2)}\pars{x_{n-1},\ldots,\widetilde{x}_a,\ldots,\widetilde{x}_b,\ldots,x_1} 
    \nonumber \\
    &+ \frac{1}{8} \sum_{\substack{(i>j)\in[1,n-1]\setminus\{e\}\\ (a,b,c)={\rm ord}(i,j,e)}} \gamma_{\braces{i,j,n}}\,p^{(+)}\pars{x_{n-1},\ldots,x_{a+1}}\,p^{(2)}\pars{x_{a-1},\ldots,\widetilde{x}_b,\ldots,\widetilde{x}_c,\ldots,x_1}
\end{align}

At every step there are $n-1$ terms $p^{(1)}(x_{n-1},\ldots,\widetilde{x}_e,\ldots,x_1)$ to compute, one for each value of the elided variable $e\in{1,n-1}$:
\begin{itemize}
    \item Each term consists of a sum of size $\sumbinom{n-2}{2}\approx n^2/2$.
    \item A total of at most $4(n-1)\sumbinom{n-2}{2} \approx O(n^3)$ computing operations. 
\end{itemize}

In the process of generating a sample of size $M$ we will need:
\begin{itemize}
    \item A tableau of total size $\binom{M}{2}$ terms
    \item The longest update take a time $(M-1)\sumbinom{M-2}{2} \approx O(M^3)$
    \item The total computing time $\sum_{i=2}^M (i-1)\sumbinom{i-2}{2} \approx O(M^4)$ operations
\end{itemize}

\subsubsection{Marginal with two elided variables}
The approximated marginals $p^{(2)}\pars{x_{n-1},\ldots,\widetilde{x}_e,\ldots,\widetilde{x}_d,\ldots,x_1}$
with all variables between $x_{n}$ and $x_{1}$ being present except for the two elided variable
$x_d$ and $x_e$ with indices in $[1,n-1]$ can be computed from the following update rule:

\begin{align} \label{eq:probdoubleelidedade}
    &p^{(2)}(x_n,\ldots,\widetilde{x}_{e},\ldots,\widetilde{x}_{d},\ldots,x_1)\nonumber\\ &= \frac{1}{2} \pars{1+\gamma_{\braces{n}}}\, p^{(2)}(x_{n-1},\ldots,\widetilde{x}_e,\ldots,\widetilde{x}_d,\ldots,x_1) \nonumber\\ 
    &+ \frac{1}{4} \sum_{\substack{i\in[1,n-1]\setminus\{e,d\}\\ (a,b,c)={\rm ord}(i,e,d)}}  \gamma_{\braces{i,n}}\,p^{(+)}\pars{x_{n-1},\ldots,x_{a+1}}\,p^{(2)}\pars{x_{a-1},\ldots,\widetilde{x}_b,\ldots,\widetilde{x}_c,\ldots,x_1}
\end{align}

At every step there are $\binom{n-1}{2}$ terms $p^{(2)}(x_n,\ldots,\widetilde{x}_{e},\ldots,\widetilde{x}_{d},\ldots,x_1)$ to compute, as we have two elided variables $\{e,d\}\in[1,n-1]$:
\begin{itemize}
    \item Each term consists of a sum of size $n-3$.
    \item A total of $3\binom{n-1}{2}(n-3) \approx O(n^3)$ computing operations. 
\end{itemize}

In the process of generating a sample of size $M$ we will need:
\begin{itemize}
    \item a tableau of total size $\binom{M}{3}$ terms
    \item the longest update take a time $\binom{M-1}{2}(M-3) \approx O(M^3)$
    \item the total computing time $\sum_{i=3}^M \binom{i-1}{2}(i-3) \approx O(M^4)$ operations
\end{itemize}

\subsubsection{Total cost estimates}

In the following table the summarize the information presented above related to the cost of storing and computing the approximated marginals $p^{(+)}$, $p^{(1)}$, $p^{(2)}$ and the sampling distribution $p$ for an $n$th step of the chain rule and the generation of a single sample of size $M$. Finally, we also give the order on the expansion in terms of cumulants for each one. 

\begin{center}
\begin{tabular}{ |c|c|c|c|c| } 
 \hline
Aux. Marg. & $p^{(+)}$ & $p^{(1)}$ & $p^{(2)}$ & $p$ \\ 
\hline
 One step &  &  & & \\ 
\hline
Terms & $O(n)$ & $O(n)$ & $O(n^2)$ & 1 \\ 
Time per term & $O(n^2)$ & $O(n^2)$ & $O(n)$ & $O(n^4)$ \\ 
Time & $O(n^3)$ & $O(n^3)$ & $O(n^3)$ &  - \\  
\hline
Per sample &  &  & & \\ 
\hline
Memory & $O(M^2)$ & $O(M^2)$ & $O(M^3)$ & $O(M)$ \\ 
Longest update & $O(M^3)$ & $O(M^3)$ & $O(M^3)$ & $O(M^4)$ \\ 
Total time & $O(M^4)$ & $O(M^4)$ & $O(M^4)$ & $O(M^5)$ \\ 
\hline
Exp. Order & 3 & 3 & 2 & 5  \\
\hline
\end{tabular}
\end{center}

In order to retain a time scaling of $O(M^5)$, we needed to truncate the cumulant expansions by at least one order for $p^{(+)}$ and $p^{(1)}$, and at least two orders for $p^{(2)}$ as there is a factor of $O(M)$ more values to compute. In practice, for the Jiuzhang experiments, we found that truncating the expansions for $p^{(+)}$, $p^{(1)}$ and $p^{(2)}$ by an additional order barely degraded the quality of the generated samples, which is why the total time complexity for computing each of the respective tables is $O(M^4)$ rather than $O(M^5)$. This choice also dramatically improves the sample generation rate with our implementation.

\section{Correlators and cumulants of GBS with binary outcomes \label{app:corrs}}

Before we derive the correlators of GBS with binary outcomes, we will do as a warm-up the qubit case with the density matrix $\hat{\varrho}$. It is easy to see that the action of the $\hat{Z}$ gate on the computational basis, i.e. $\hat{Z}\ket{x}=(-1)^x\ket{x}$ is characterized by the single-bit parity function. One can generalize this and define the operator $\hat{Z}_A$ such that $\hat{Z}_A\ket{\bar{x}}=(-1)^{\sum_{k\in A}x_k}\ket{\bar{x}}$, where
\begin{equation}
    \hat{Z}_A=\left(\bigotimes_{i\in A} \hat{Z}_i\right)\otimes \hat{\mathbb{I}}_{\bar{A}},
\end{equation}
and $\hat{\mathbb{I}}_{\bar{A}}$ is the identity operator over the complement set to $A$. It easy to see that
\begin{equation}
    \text{Tr}\left[\hat{Z}_A\hat{\varrho}\right]=\text{Tr}\left[\hat{\varrho}_A\bigotimes_{i\in A} \hat{Z}_i\right]=
    \frac{1}{2^{|A|}}\sum_{\bar{x}\in\{0,1\}^{|A|}}p(\bar{x}_A)(-1)^{\sum_{k\in A}x_k},
\end{equation}
where $p(\bar{x}_A)$ is the probability of outcome $\bar{x}_A$ if we measure $\hat{\varrho}_A$ in the computational basis, i.e. the correlator corresponding to the probability distribution resulting from measuring the quantum state $\hat{\varrho}$ into the computational basis.

\subsection{Correlators of GBS with binary outcomes\label{app:corr_gbs}}
The outcome of a threshold detector measurement of a multimode Gaussian state, $\hat{\varrho}$, is a bitstring $\bar{x}=(x_1, \dots, x_M)$ with $M$ the number of modes in the system. The outcome $x_k = 0$ indicates that no photons have been detected at mode $k$, while $x_k = 1$ indicates that one or more photons arrived at the detector (i.e. that the detector ``clicked''). For each mode, this measurement is represented by the two-element POVM $\{|0_k\rangle\langle 0_k|, \hat{\mathbb{I}}_k - |0_k\rangle\langle 0_k|\}$, where $\hat{\mathbb{I}}_k$ is the identity operator at mode $k$ and $|0_k\rangle$ the vacuum state in the same mode. 
Similarly as for qubits we have $Z=2|0\rangle\langle 0|-\hat{\mathbb{I}}$, one can define a generalized version of the $Z$ operator to the Fock basis as $Z=2|0_k\rangle\langle 0_k|-\hat{\mathbb{I}}_k$, which after some derivation shown below, leads to
\begin{equation}
    c(S) =\mathrm{Tr}\left[\hat{\varrho}\left(\bigotimes_{k\in S}[2|0_k\rangle\langle 0_k|-\hat{\mathbb{I}}_k]\right)\otimes \hat{\mathbb{I}}_{[M]\setminus S}\right],
    \label{eq:correlator_operator2}
\end{equation}
which can be further developed as a sum of Gaussian integrals as we detail below.

\subsubsection{Formal Derivation}

Given a multimode measurement outcome $\bar{x}$, let $\mathcal{C}$ be the subset of $[M]=\{1, \dots, M\}$ for which $x_k=1$ for all $k\in \mathcal{C}$. Then, we can associate $\bar{x}$ to the measurement of the operator
\begin{equation}
    \hat{\Pi}(\bar{x}) = \bigotimes_{k\in \mathcal{C}}(\hat{\mathbb{I}}_k - |0_k\rangle\langle0_k|)\bigotimes_{j\in [M]\setminus\mathcal{C}}|0_k\rangle\langle0_k|.
    \label{eq:threshold_measurement_oper}
\end{equation}
Alternatively, we can write this operator as 
\begin{equation}
    \hat{\Pi}(\bar{x}) = \bigotimes_{k=1}^M (\hat{\mathbb{I}}_k-|0_k\rangle\langle 0_k|)^{x_k}(|0_k\rangle\langle0_k|)^{1-x_k}.
    \label{eq:threshold_measurement_oper_2}
\end{equation}
From these relations, one can readily see that $\sum_{\bar{x}\in \{0,1\}^M}\hat{\Pi}(\bar{x})=\hat{\mathbb{I}}$, with $\hat{\mathbb{I}}$ the identity operator over the $M$-mode Hilbert space of the system.

The probability of obtaining the outcome $\bar{x}$ when measuring the system in the state $\hat{\varrho}$ is given by 
\begin{equation}
    p(\bar{x})=\mathrm{Tr}[\hat{\Pi}(\bar{x})\hat{\varrho}].
    \label{eq:prob_distribution_operator}
\end{equation}
When combined with the definition of correlator in Eq.~\eqref{eq:correlators_def_bitstrings} , we can use this expression to associate a Hermitian operator to each correlator $c(S)$. Indeed, using the notation
$\hat{\mathbb{I}}_A$ for the identity operator over the Hilbert space associated to the modes in $A$, we have
\begin{align}
    c(S) &= \sum_{\bar{x}\in\{0,1\}^M}p(\bar{x})\left(\prod_{k\in S}(-1)^{x_k}\right)\nonumber\\
    &= \mathrm{Tr}\left[\hat{\varrho}\,\sum_{\bar{x}\in \{0,1\}^M}\left(\prod_{k\in S}(-1)^{x_k}\right)\left(\bigotimes_{l=1}^M (\hat{\mathbb{I}}_l-|0_l\rangle\langle 0_l|)^{x_l}(|0_l\rangle\langle 0_l|)^{1-x_l}\right)\right] \nonumber\\
    &= \mathrm{Tr}\left[\hat{\varrho}\bigotimes_{l=1}^M \sum_{x_l\in \{0,1\}}\left((-1)^{x_{l\in S}}(\hat{\mathbb{I}}_l-|0_l\rangle\langle 0_l|)^{x_l}(|0_l\rangle\langle 0_l|)^{1-x_l}\right)\right] \nonumber\\
    \label{eq:correlator_operator}
\end{align}
where $(-1)^{x_{l\in S}}=(-1)^{x_l}$ if $l\in S$ or $1$ otherwise. 
The sum over both values of $x_l$ contribute a term $\hat{\mathbb{I}}_l$ when $l\notin S$ and $2|0_l\rangle\langle 0_l|-\hat{\mathbb{I}}_l$ when $l\in S$, leading to
\begin{equation}
    c(S) =\mathrm{Tr}\left[\hat{\varrho}\left(\bigotimes_{k\in S}[2|0_k\rangle\langle 0_k|-\hat{\mathbb{I}}_k]\right)\otimes \hat{\mathbb{I}}_{[M]\setminus S}\right],
    \label{eq:correlator_operator2_app}
\end{equation}
 Note that the operators $\{2|0_k\rangle\langle0_k| -\hat{\mathbb{I}}_k, \hat{\mathbb{I}}_k\}_{k=1}^M$ do not form a POVM.

Taking into account that 
\begin{equation}
    \bigotimes_{k\in S}[2|0_k\rangle\langle 0_k|-\hat{\mathbb{I}}_k]=\sum_{R\subseteq S}\left[\bigotimes_{k\in R}(2|0_k\rangle\langle 0_k|)\bigotimes_{\ell\in S\setminus R}(-\hat{\mathbb{I}}_\ell)\right],
    \label{eq:corr_oper_reexpressed}
\end{equation}
we can recast $c(S)$ as
\begin{equation}
    c(S) = (-1)^{|S|}\sum_{R\subseteq S}(-2)^{|R|}\langle 0 _R|\hat{\varrho}_R|0_R\rangle,
    \label{eq:correlator_sum_of_expectation}
\end{equation}
where $|0_R\rangle = \bigotimes_{k\in R}|0_k\rangle$, and $\hat{\varrho}_R$ is the reduced state obtained by tracing out all the modes that are not in $R$, i.e. $\hat{\varrho}_R = \mathrm{Tr}_{[M]\setminus R}(\hat{\varrho})$.

The Gaussian state $\hat{\varrho}$ can be completely parametrized by a $2M\times2M$ covariance matrix $\boldsymbol{\sigma}$ and a vector of first moments $\boldsymbol{\mu}$ of length $2M$~\cite{serafini2023quantum}. The reduced state $\hat{\varrho}_R$ is also Gaussian, and can be parametrized by a $2|R|\times2|R|$ covariance matrix $\sigma_R$ and a vector of means $\boldsymbol{\mu}_R$, which can be obtained from $\boldsymbol{\sigma}$ and $\boldsymbol{\mu}$ according to the following prescription: $\boldsymbol{\mu}_R$ is obtained by removing the $k$th and $(k+M)$th elements of $\boldsymbol{\mu}$ for $k\in [M]\setminus R$, while $\boldsymbol{\sigma}_R$ is obtained by removing the $k$th and $(k+M)$th rows and columns of $\boldsymbol{\sigma}$ for $k\in [M]\setminus R$~\cite{serafini2023quantum}.

In terms of $\boldsymbol{\sigma}_R$ and $\boldsymbol{\mu}_R$, it can be shown that~\cite{hamilton2017gaussian, kruse2019detailed}
\begin{equation}
    \langle 0_R|\hat{\varrho}_R|0_R\rangle = \frac{\exp\left[-\frac{1}{2}\boldsymbol{\mu}_R^{\mathrm{T}}\left(\boldsymbol{\sigma}_R + \frac{\hbar}{2}\mathbb{I}\right)^{-1}\boldsymbol{\mu}_R\right]}{\sqrt{\mathrm{det}\left[\frac{1}{\hbar}\left(\boldsymbol{\sigma}_R + \frac{\hbar}{2}\mathbb{I}\right)\right]}},
    \label{eq:vacuum_gaussian_prob}
\end{equation}
with $\mathbb{I}$ the $2|R|\times2|R|$ identity matrix. For the non-displaced ($\boldsymbol{\mu}=0$) case of the Jiuzhang experiments, the correlators $c(S)$ take the following form as a function of $\boldsymbol{\sigma}$:
\begin{equation}
    c(S) = (-1)^{|S|}\sum_{R\subseteq S}\frac{(-2)^{|R|}}{\sqrt{\mathrm{det}\left[\frac{1}{\hbar}\left(\boldsymbol{\sigma}_R + \frac{\hbar}{2}\mathbb{I}\right)\right]}}.
    \label{eq:correlator_sum_cov}
\end{equation}

\subsection{cumulant expression\label{app:cumul_exp}}

The cumulants $\kappa$ required in the sampling procedure are computed from the correlators $c(S)$ for all sets $S$ of size 1 to $K$ composed of elements of $[M]$ using the relation
\begin{equation}
\kappa(S)= \sum_{\pi \in \mathcal{P}(S)}(-1)^{|\pi|-1}(|\pi|-1)!\prod_{B\in \pi}c(B).
\end{equation}

\section{Comparison to other implementations and experiment \label{app:comparison}}

Here we discuss the performance and scalability of the algorithm for simulating GBS presented in Ref.~\cite{oh2024classical} (which we shall refer to as `the Chicago method'). There are two main stages of the algorithm: formation of the matrix product state (MPS) which is costly, but need only be done once, and sampling, which is less expensive but must of course be repeated every time we generate a sample. The MPS represents the output state of some quantum circuit under the influence of photon loss. It is by taking the effect of photon loss into account that the Chicago method achieves advantage over prior algorithms. The idea behind the approach is that losses allow the output state to be written as a convex mixture of Gaussian states with significantly lower squeezing, which can be efficiently created, stored, and sampled from, however, a random set of local displacement operations need to be applied when generating each sample. Local operations have no major impact on the scalability of sampling from an MPS, the costs are mostly equivalent to those of the pure Gaussian state with weaker squeezing resulting from the convex decomposition facilitated by the presence of losses.
The prepared MPS is saved to disk, and then loaded separately for the sampling calculation.

\subsection{Time complexity\label{app:time_complexity}}

\paragraph{MPS Formation}

First we consider the asymptotic time complexity of the MPS formation stage of the Chicago method. This is reported as \(\mathcal{O}(cMd\chi^{2})\) where \(\chi\) is the bond dimension of the MPS tensors, \(d\) is the physical dimension of the local Hilbert space, i.e. the maximum number of photons per mode, \(M\) is the number of modes, and \(c\) is ``a parameter that depends on the system's characteristics''.
The bond dimension \(\chi\) is a parameter of the MPS representation itself that determines the available degrees of freedom -- the ability to compress the representation of a state by limiting \(\chi\) is at the core of tensor network methods. Crucially, for quantum states, the bond dimension is linked to the Schmidt rank of any bipartition of the system and therefore limits the amount of entanglement that can be represented by the MPS. This means that MPS (and other tensor decompositions) can be used to form a well controlled approximation of a quantum state, and the lower the entanglement in the system, the more efficiently the state can be represented. They use values of \(\chi\) ranging from 2 to 10,000, with \(\chi=10,000\) for the majority of their numerical simulations. 
The physical dimension \(d\) is simpler to understand, being the maximum number of photons that may appear in any given mode. It obviously must be truncated to some finite value, \(d=4\) was used during MPS construction, while \(d=10\) during sampling to allow for the contribution of additional photons resulting from the displacement operation.
The number of modes \(M\) is a parameter of the experiment being simulated, and ranges from 72 to 288 in the numerical results.
The final parameter, \(c\), is the complexity of computing a hafnian, which is exponential in the mean number of photons of each mode's thermal state, and consequently dominated by the largest such photon number in the system. More precisely, the scaling is given in the supplementary material of Ref.~\cite{oh2024classical} as \(\widetilde{O}(2^{(\mathbf{n}_{1} + \mathbf{n}_{2})/2})\), where \(\mathbf{n}_{i}\) are photon occupation patterns appearing in expectation values $\bra{{n}_{2}}U\ket{{n}_{1}}$ leading to the tensor terms. Due to the low squeezing resulting from the convex decomposition, these occupation patterns have limited total number of photons, making the hafnian calculation more efficient than directly sampling from the output state using the best-known classical algorithm.

\paragraph{Sampling}

The serial complexity of the sampling procedure is dominated by the bond dimension of the MPS tensors, with an asymptotic complexity of \(\mathcal{O}(M\chi^{2}d)\) \emph{per sample}. Again, we note that during the sampling phase the parameter \(d\), the physical dimension of the MPS tensor which limits the number of photons allowed per site, is increased from 4 to 10. This is to accommodate an operation in which photons in the prepared state are randomly displaced immediately before each sample is taken, and which may cause the photon number to increase on particular sites. In practical terms there is more work to do in the sampling procedure than the MPS preparation procedure whenever the number of samples \(N\) is greater than \(c\) the complexity of computing a hafnian. This is reflected in the runtimes for each stage, where constructing the MPS took around 10 minutes for the largest size problems, while generating ten million samples took around an hour, despite the parallelism available during the sampling step.

\subsection{Space complexity\label{app:space_complexity}}

The dominant data structure is the matrix product state which has a total size of \(\mathcal{O}(Md\chi^{2})\), with one tensor per mode -- once again the bond dimension, \(\chi\), is dominant. This structure is at its largest during the sampling procedure when \(d\) is increased to 10. Since each element of the tensor is a 16 byte complex double, this means each tensor requires 16 GB of memory at \(\chi=10,000\). The GPUs used in this experiment had 40 GiB of memory each, so were only capable of carrying two site tensors at once, and at least one copy is required during the sample calculation.

\subsection{Parallel scalability\label{app:parallel_scalability}}

Due to the large size of the MPS tensors the principle bottleneck in scaling the code is the available memory on each GPU. In both the MPS preparation procedure, and the sampling procedure, only one site tensor is loaded per GPU meaning for a system with \(M\) modes, exactly \(M\) GPUs are used. Both stages do take advantage of the shared memory thread-parallelism available on each GPU, with multiple hafnians calculated in parallel during the MPS construction phase, and multiple samples calculated in parallel during the sampling phase.
This parallelism is limited in both cases by the memory available on the GPU -- in the sampling stage for the largest system the authors fix the number of samples calculated in parallel to 100.

The MPS construction phase benefits from requiring no communication between GPUs, whereas the sampling stage essentially requires a full contraction of the network so vectors of length \(\chi\) must be passed from GPU to GPU to calculate the overall sample result. This step can however be pipelined, with a number of steps equal to \(M+\frac{N}{B}-1\) where \(M\) is the number of modes and thus GPUs and pipeline stages, \(N\) is the total number of samples being calculated, and \(B\) is the number of samples calculated in parallel on each GPU so there are \(M\) timesteps before any result is produced, and then \(\frac{N}{B}\) results produced at each subsequent timestep. The complexity of each step in the pipeline is, as described above, \(\mathcal{O}(\chi^{2}d)\).

Due to the limitation of one GPU per mode, there is no scalability in terms of number of GPUs, but runtime would nevertheless be improved by GPUs with more memory, which would allow more hafnians to be calculated in parallel giving a linear improvement to the MPS construction procedure, or more samples to be calculated in parallel (\(B\)) on each GPU in the sampling stage. 
The biggest disadvantage in scaling the simulation to larger GBS experiments is the need for one GPU per mode, making the simulation of few thousand modes,
like in the recent Jiuzhang 4.0 \cite{liu2025} experiment, very costly to implement.

\section{Benchmarking theory and implementation \label{app:benchmark}}

In this section, we explain the different benchmarking tools used in the validation of our sampler. For all these test, we compared sample metrics with those predicted by the ground truth of the Jiuzhang experiments, which in turn can be obtained using the data that has been made publicly available by the authors of the experiments~\cite{ustc2020experimental, ustc2021raw, ustc2023raw}. Before explaining the theory and purpose of the different validation methods, in the following, we will explain how to obtain the ground truth that each of these tests uses.

The Jiuzhang 2.0 experiment consists of a set of $k=25$ two-mode squeezed states with different squeezing parameters that are sent into a lossy linear interferometer with $M=144$ output ports. The light exiting the interferometer is measured using threshold detectors. The authors of the experiment made available a list of of squeezing parameters $\boldsymbol{r}=(r_1, \dots, r_{k})$ and a $2k \times M$ complex transmission matrix $\boldsymbol{T}$ representing the action of the interferometer. The ground truth distribution of the experiment corresponds to the probability distribution of the threshold detection outcomes, $p(\bar{x})$, that are obtained after the measurement of the Gaussian state exiting the interferometer. This distribution is completely parametrized by the quadrature covariance matrix of the output state, $\boldsymbol{\sigma}_{\mathrm{GT}}$, which can be obtained from $\boldsymbol{r}$ and $\boldsymbol{T}$ as follows.

The $k$ input two-mode squeezed states form a $2k$-mode non-displaced Gaussian state that is completely parametrized by a $4k\times 4k$ covariance matrix
\begin{equation}
    \boldsymbol{\sigma}_{\mathrm{IN}} = \frac{\hbar}{2}\left[\bigoplus_{j=1}^{k}\begin{pmatrix}\cosh(r_j) & \sinh(r_j)\\\sinh(r_j) & \cosh(r_j)\end{pmatrix}\right]\oplus\left[\bigoplus_{j=1}^{k}\begin{pmatrix}\cosh(r_j) & -\sinh(r_j)\\-\sinh(r_j) & \cosh(r_j)\end{pmatrix}\right].
    \label{eq:jiuzhang_input_cov}
\end{equation}
By defining 
\begin{equation}
    \boldsymbol{V} = \begin{pmatrix}
        \mathrm{Re}(\boldsymbol{T}) & -\mathrm{Im}(\boldsymbol{T})\\
        \mathrm{Im}(\boldsymbol{T}) & \mathrm{Re}(\boldsymbol{T})
    \end{pmatrix},
    \label{eq:symplectic_inter}
\end{equation}
the $2M\times2M$ ground truth covariance matrix can be computed as
\begin{equation}
    \boldsymbol{\sigma}_{\mathrm{GT}}=\boldsymbol{V}\boldsymbol{\sigma}_{\mathrm{IN}}\boldsymbol{V}^{\mathrm{T}} + \frac{\hbar}{2}\left(\mathbb{I} - \boldsymbol{V}\boldsymbol{V}^{\mathrm{T}}\right).
    \label{eq:ground_truth_cov}
\end{equation}

Given a measurement outcome $\bar{x}$, the corresponding ground truth probability is computed as 
\begin{equation}
    p(\bar{x}) = \frac{\mathrm{tor}[\boldsymbol{O}_{(\bar{x})}]}{\sqrt{\mathrm{det}(\boldsymbol{\Sigma}_{\text{GT}})}},
    \label{eq:ground_truth_prob_tor}
\end{equation}
where $\boldsymbol{O}=\mathbb{I}-\boldsymbol{\Sigma}^{-1}_{\text{GT}}$, $\boldsymbol{\Sigma}_{\text{GT}} = \frac{1}{2}\mathbb{I} + \frac{1}{\hbar}\boldsymbol{R}\boldsymbol{\sigma}_{\text{GT}}\boldsymbol{R}^\dagger$, and $\boldsymbol{R} = \frac{1}{\sqrt{2}}\big(\begin{smallmatrix}\mathbb{I}&i\mathbb{I}\\\mathbb{I}&-i\mathbb{I}\end{smallmatrix}\big)$. The matrix $\boldsymbol{O}_{(\bar{x})}$ is obtained by removing the $k$th and $(k+M)$th rows and columns from $\boldsymbol{O}$ whenever $x_k = 0$. Finally, $\mathrm{tor}(\boldsymbol{M})$ is the ``Torontonian''~\cite{quesada2018gaussian} of the $2n\times2n$ matrix $\boldsymbol{M}$, and it is computed as
\begin{equation}
    \mathrm{tor}(\boldsymbol{M}) = \sum_{R\subseteq[n]}\frac{(-1)^{|R|}}{\sqrt{\mathrm{det}(\mathbb{I}-\boldsymbol{M}_R)}},
    \label{eq:torontonian_def}
\end{equation}
where, again, $\boldsymbol{M}_R$ is constructed from $\boldsymbol{M}$ by keeping only its $k$th and $(k+n)$th rows and columns with $k\in R$.

\subsection{\label{app:cumulants_benchmark}Click cumulants, Pearson and Spearman coefficients}

The comparison of click or photon-number cumulants has been used in the validation of all the GBS experiments implemented so far. This test relies on the idea that non-trivial high-order correlations of the GBS distribution are a feature that cannot be easily reproduced by classical strategies. Thus, by demonstrating that the correlations in the experimental samples match those predicted by the ``ground truth'' of the experiment better than any other classical model or sampler, we can build confidence in the correct functioning of the boson sampler. 

Mathematically, these correlations are represented by cumulants, which are defined in terms of the joint moments of a set of random variables $(X_1, \dots, X_M)$ as
\begin{equation}
    \tilde{\kappa}(X_{j_1}, \dots, X_{j_n})= \sum_{\pi \in \mathcal{P}([n])}(-1)^{|\pi|-1}(|\pi|-1)!\prod_{b\in \pi}\left\langle\prod_{k\in b}X_{j_k}\right\rangle,
    \label{eq:cumulants_test_def}
\end{equation}
where $\{j_1, \dots,j_n\} \subseteq [M] = \{1, \dots, M\}$ is a set of indices, and $n \leq M$ is the order of the cumulant. The sum runs over the collection of all partitions, $\pi$, of $[n]=\{1,\dots, n\}$. The index $b$ represents each block of the partition, and $|\pi|$ stands for the number of blocks in the partition. From this definition, one can readily see that the first-order cumulants are the expectation values of the random variables $\tilde{\kappa}(X_k)=\langle X_k\rangle$, while the second-order cumulants correspond to the covariances of these variables $\tilde{\kappa}(X_j,X_k)=\langle X_jX_k\rangle-\langle X_j\rangle\langle X_k\rangle$.

When estimating cumulants from a given set of samples, it is possible to use a very similar expression to Eq.~\eqref{eq:cumulants_test_def}. We need only replace $\left\langle\prod_{k\in b}X_{j_k}\right\rangle$ by the corresponding estimator of the mean, and modify the weights $w(\pi)=(-1)^{|\pi|-1}(|\pi|-1)!$ by a factor depending on the number of samples used. We refer the reader to Ref.~\cite{smith2020tutorial} for details on the definition of the unbiased estimator of the cumulants.

For the Jiuzhang experiments, the random variables are associated to the detection of a ``click'' (i.e. one or more photons) in the threshold detector measurement, $X_k \Leftrightarrow \hat{\mathbb{I}}_k - |0_k\rangle \langle 0_k|$. On this account, the joint moments of these variables can be computed as
\begin{equation}
    \left\langle\prod_{k\in b}X_{j_k}\right\rangle = \mathrm{Tr}\left[\hat{\varrho}\,\bigotimes_{k\in b}\left(\hat{\mathbb{I}}_{j_k} - |0_{j_k}\rangle\langle0_{j_k}|\right)\right] = p(x_{j_k} = 1 \text{ for } k\in b).
    \label{eq:joint_moments_threshold}
\end{equation}
That is, the theoretical joint moments can be computed as marginals of the GBS distributions. It is worth noticing that Eq.~\eqref{eq:cumulants_test_def} is, up to a factor of $(-2)^n$ (and a shift for $n=1$), equivalent to the expression of $\kappa(S)$ shown in Eq.~\eqref{eq:cumulant_exponential}.

In principle, all click cumulants can be estimated if we have enough samples so that the statistical uncertainty is at least one order of magnitude lower than its ground truth value. The computation has only linear scaling with the order. 
However, for some distributions as the one we face here, the cumulant values decrease exponentially with the order, making their estimation inefficient beyond a given threshold value. 
In addition, their number being given by the binomial $\binom{M}{d}$, increasing
polynomially in the number of modes $M$ and exponentially with the order $d$ (until it reached $M/2$ where the numbers start decreasing again), make it almost impossible to
collect all data or compute averages beyond a handful of orders. 
So far, all previous cumulant analyses of samples from experimental or classical simulation 
have computed full cumulants up to third order, and perform a random selection of cumulants for fourth and fifth order. In this work, after some improvement on the implementation of cumulant computation with a GPU, we were able to analyse all cumulants up to fourth order, but we are still restricted to a random sub-ensemble at fifth order, namely a subset of size $1/24$ the total number of cumulants of fifth order.

\subsubsection{Scatter plots}

A first tool to carry out the comparison of click cumulants is a scatter plot for each cumulant order, where one of the variables (the $x$-axis variable) corresponds to the ground truth cumulants, and the second variable (the $y$-axis variable) corresponds to the estimated cumulants from either the experimental samples or from samples generated from a classical algorithm. We show this type of plots in Figs.~\ref{fig:validation_J2_2}a,~\ref{fig:validation_J25}a, and~\ref{fig:validation_J33}a. If the estimated cumulants were precisely those predicted by the ground truth, all the dots in the scatter plot would lie on a straight line with unit slope and null intercept with the $y$-axis. On this account, estimating the parameters of a linear fit between sample and theoretical cumulants serves as a measure of how correlated a set of samples is to the ground truth of the experiments. For the validation of our sampler, we estimated the cumulants up to fifth order using a total of $10^7$ samples from both the experiments and the double-elision algorithm. We computed the corresponding linear fits (see Figs.~\ref{fig:validation_J2_2}a,~\ref{fig:validation_J25}a, and~\ref{fig:validation_J33}a) between theoretical and sample cumulants, and we found that our sampler generally obtains slopes closer to the ideal value than those of the experimental cumulants.

\subsubsection{Pearson or Spearman}

To further demonstrate that a set of estimated cumulants matches those predicted by the ground truth of the experiment, one usually shows that the Pearson or Spearman correlation coefficients between the two sets of data are close to 1 for different orders of the cumulants. This reveals the presence of a linear or monotonic relation between sample and ground truth correlations. More precisely, for each order $n$, let $\{\tilde{\kappa}_T^{(k)}\}_{k=1}^L$ and $\{\tilde{\kappa}_S^{(k)}\}_{k=1}^L$ be two sets of $L$ ground truth and sample cumulants, respectively. The Pearson correlation coefficient is computed as
\begin{equation}
    r = \frac{\sum_{k=1}^L(\tilde{\kappa}_S^{(k)} - \langle\tilde{\kappa}_S\rangle)(\tilde{\kappa}_T^{(k)} - \langle\tilde{\kappa}_T\rangle)}{\sqrt{\sum_{k=1}^L(\tilde{\kappa}_S^{(k)} - \langle\tilde{\kappa}_S\rangle)^2\sum_{k=1}^L(\tilde{\kappa}_T^{(k)} - \langle\tilde{\kappa}_T\rangle)^2}},
    \label{eq:pearson_correlation}
\end{equation}
where $\langle\tilde{\kappa}_S\rangle = (1/L)\sum_{k=1}^L\tilde{\kappa}_S^{(k)}$ and $\langle\tilde{\kappa}_T\rangle$ is similarly defined. The Spearman correlation coefficient is computed using the same expression in Eq.~\eqref{eq:pearson_correlation}, but replacing $
\tilde{\kappa}_T^{(k)}$, $\tilde{\kappa}_S^{(k)}$ by their corresponding ``rank'' (i.e. by their positions when the sets $\{\tilde{\kappa}_T^{(k)}\}_{k=1}^L$ and $\{\tilde{\kappa}_S^{(k)}\}_{k=1}^L$ are ordered).

When comparing to classical models or samplers, the goal is to demonstrate that the correlation coefficients between the ground truth and classical cumulants are not as high as those between ground truth and experimental cumulants. For the validation of our sampler, we computed the Pearson correlation coefficients between sample and ground truth cumulants up to fifth order. For this computation, we used $10^7$ samples from both the experiment and the double-elision algorithm (see Figs.~\ref{fig:validation_J2_2}b,~\ref{fig:validation_J25}b, and~\ref{fig:validation_J33}b). We find that our sampler obtains higher values of the Pearson correlation coefficient than the experimental samples for all orders.

\subsection{\label{app:total_clicks}Total click distribution}
The computation and estimation of the probability of a single experimental sample $\bar{x}=(x_1, \dots, x_M)$ with respect to the ground truth distribution of a GBS experiment are considered to be computationally hard tasks. The cost of computing these probabilities increases exponentially with the number of clicks in the experimental samples, while their estimation requires exponentially-many of these samples. Consequently, one cannot readily validate a GBS experiment by directly estimating its corresponding probability distribution and comparing it with that predicted by the associated theoretical model. One can partially overcome this limitation by computing and estimating probabilities of ``coarse-grained'' detection events instead of dealing with individual, ``fine-grained'' experimental samples. 

A coarse-grained detection event is defined by a grouping of the detection modes, which is mathematically defined by a partition $\Lambda = \{\Lambda_1, \dots, \Lambda_\ell\}$ of $[M] = \{1, \dots, M\}$, and a coarse grained detection pattern $\bar{C} = (C_1, \dots, C_\ell)$. The corresponding probability is computed as 
\begin{equation}
    p(\bar{C}, \Lambda) =\sum_{\bar{x} \text{ comp. } (\bar{C}, \Lambda)} p(\bar{x}),
    \label{eq:coarse_grained_distribution}
\end{equation}
where the sum is taken over all detection patterns $\bar{x}$ that are ``compatible'' with $\bar{C}$, that is, those satisfying $\sum_{j\in\Lambda_k}x_j = C_k$ for all $k\in \{1, \dots, \ell\}$. A particularly important type of the coarse-grained distribution is the ``total click distribution'', $p(C)$, obtained when the partition $\Lambda$ has only one block: 
\begin{equation}
    p(C) =\sum_{|\bar{x}| = C } p(\bar{x}),
    \label{eq:total_click_distribution}
\end{equation}
where, let us recall, $|\bar{x}| = \sum_{k=1}^M x_k$~\cite{bulmer2024simulating}.

From Eqs.~\eqref{eq:coarse_grained_distribution} and~\eqref{eq:total_click_distribution}, it appears that the computation of coarse-grained probabilities is \textit{harder} than obtaining the probability of a single detection event. However, it has been shown~\cite{drummond2022simulating} that $p(\bar{C})$ can be efficiently estimated when the number of blocks, $|\Lambda|$, is not too big. In particular, the theoretical value of $p(C)$ can be efficiently estimated using phase-space techniques, and it can be experimentally estimated using polynomially-many samples. We refer the reader to Ref.~\cite{drummond2022simulating} for details on the definition of the numerical phase-space methods for computing the total click distribution.

The partial validation of GBS experiments using coarse-grained detection events relies on the idea that a classical model or sampler trying to explain or simulate the experiment should be able to reproduce the coarse-grained distributions of the theoretical model of the implementation. If the experimentally estimated coarse-grained probabilities are closer to those predicted by the ground truth than those obtained by classical samplers or models, we gain more trust in the correct functioning of the boson sampler. The Jiuzhang experiments have been partially validated using the comparison of total click distributions. This method has been able to rule out classical models such as thermal, coherent and squashed states models, and other classical samplers.

For the validation of our sampler, we estimated the double-elision total click distribution, as well as that of the experimental outcomes, using a total of $10^7$ samples. The ground truth distributions were computed using a \texttt{Python} implementation of the phase-space technique for estimating the total click distribution available at the \texttt{thewalrus} library~\cite{gupt2019walrus} (namely, the function \texttt{grouped\_click\_probabilities}). As shown in Figs.~\ref{fig:validation_J25}c, and~\ref{fig:validation_J33}c, the total click distribution of our sampler better aligns with that predicted by the ground truth.

\subsection{\label{app:xeb} Cross-entropy benchmarking (XEB)}
A natural way of making sure that an experimental implementation of GBS is following its intended theoretical model, is to verify how similar the probability distribution of the experimental samples is to the ground truth distribution. One of the most useful statistical measures that allow to check this similarity is the total variation distance. However, it has been shown that the accurate estimation of these quantities in the context of GBS requires a number of experimental samples that grows exponentially with the number of modes in the system, making this validation strategy inefficient. There exist other ``sample-efficient'' measures of similarity that act as proxies of the total variation distance without being as decisive. The XEB score is one of these measures, and has been widely used in the validation of real-world implementations of qubit-based quantum random circuit sampling \cite{Arute_2019}.

Considering a GBS implementation using threshold detectors, let $\{\bar{x}_k\}_{k=1}^{N}$ be a set of experimental samples sharing the same total number of clicks, $C$. The XEB score for this collection of samples is defined as
\begin{equation}
    \mathrm{XEB}(C) = \frac{1}{N}\sum_{k=1}^N \log\left[\binom{M}{C}\frac{p(\bar{x}_k)}{p(C)}\right],
    \label{eq:xeb_definition}
\end{equation}
where $p(\bar{x})$ is the ground truth probability distribution, and $p(C)$ is the ground truth total click distribution. Note that by depending on the calculation of probabilities of individual samples, XEB is a \textit{computationally inefficient} validation measure. This means that while it can be estimated using polynomially-many samples, one is restricted to compute it for small values of the total number of clicks, typically $C<27$. This makes the XEB verification of bright GBS experiments (that lead to samples with many clicks) particularly challenging.

The main idea behind the XEB validation of GBS is to compute $\mathrm{XEB}(C)$ as a function of $C$ for a set of experimental samples, and for samples obtained using classical strategies. By showing that the experimental samples generally obtain higher XEB scores, we obtain more confidence in the possibility that the boson sampler follows the ground truth more closely than the classically generated samples. Intuitively, this is due to the fact that high XEB scores are obtained for generally high values of $p(\bar{x}_k)$, which, in principle, are obtained by samples that follow a probability distribution close to the ground truth.

It is worth mentioning that there is no evidence that obtaining a high XEB score implies that the distribution of the corresponding samples is close in total variation distance to the ground truth. For instance, there are ways of spoofing the XEB test by sampling from a reference distribution that captures some of the features of the ground truth~\cite{oh2023spoofing}. 

For the validation of our sampler in the low-brightness regime, which corresponds to total click distributions that are supported mostly in the range $0\leq C< 27$, we defined sets of $4000$ experimental and double-elision samples for each value of $C$. Using these samples, we estimated the values of $\text{XEB}(C)$ and plotted the results as functions of $C$. For the J2-P65-2 experiment in particular (see Figure~\ref{fig:validation_J2_2}(c)), we found that our sampler obtains XEB scores that are as high as those obtained by the experimental samples, suggesting that the distribution of the sampler is as similar to the ground truth distribution as that of the experimental samples.

\end{document}